\documentclass[a4paper,12pt]{article}
\usepackage{jheppub}
                     
\usepackage{amsmath,amssymb,amsthm,amscd,dsfont,graphicx,psfrag}
\input epsf.sty

\DeclareMathOperator{\su2}{SU(2)}
\DeclareMathOperator{\suN}{SU(N)}

\DeclareMathOperator{\o2}{O(2)}

\let\originalleft\left
\let\originalright\right
\renewcommand{\left}{\mathopen{}\mathclose\bgroup\originalleft}
\renewcommand{\right}{\aftergroup\egroup\originalright}

\newcommand{\CT}{{\cal T}}

\def\IN{{\mathbb N}}
\def\IZ{{\mathbb Z}}
\def\IR{{\mathbb R}}
\def\IC{{\mathbb C}}

\def\IF{{\mathbb F}}

\newcommand{\re}{{\rm e}}
\newcommand{\x}{{\rm x}}
\newcommand{\y}{{\rm y}}

\newcommand{\ri}{{\rm i}}
\newcommand{\rd}{{\rm d}}

\newcommand{\tg}{{\tt g}}

\newcommand{\be}{\begin{equation}}
\newcommand{\ee}{\end{equation}}
\newcommand{\ba}{\begin{aligned}}
\newcommand{\ea}{\end{aligned}}
\newcommand{\ben}{\begin{eqnarray}\displaystyle}
\newcommand{\een}{\end{eqnarray}}

\newdimen\tableauside\tableauside=1.0ex
\newdimen\tableaurule\tableaurule=0.4pt
\newdimen\tableaustep
\def\phantomhrule#1{\hbox{\vbox to0pt{\hrule height\tableaurule width#1\vss}}}
\def\phantomvrule#1{\vbox{\hbox to0pt{\vrule width\tableaurule height#1\hss}}}
\def\sqr{\vbox{%
  \phantomhrule\tableaustep
  \hbox{\phantomvrule\tableaustep\kern\tableaustep\phantomvrule\tableaustep}%
  \hbox{\vbox{\phantomhrule\tableauside}\kern-\tableaurule}}}
\def\squares#1{\hbox{\count0=#1\noindent\loop\sqr
  \advance\count0 by-1 \ifnum\count0>0\repeat}}
\def\tableau#1{\vcenter{\offinterlineskip
  \tableaustep=\tableauside\advance\tableaustep by-\tableaurule
  \kern\normallineskip\hbox
    {\kern\normallineskip\vbox
      {\gettableau#1 0 }%
     \kern\normallineskip\kern\tableaurule}%
  \kern\normallineskip\kern\tableaurule}}
\def\gettableau#1{\ifnum#1=0\let\next=\null\else
\squares{#1}\let\next=\gettableau\fi\next}

\preprint{CERN-TH-2023-169}

\title{\boldmath Painlev\'e Kernels and Surface Defects at Strong Coupling}

\author{Matijn Fran\c cois and Alba Grassi}
\affiliation{Section de Math\'{e}matiques, Universit\'{e} de Gen\`{e}ve, 1211, Gen\`{e}ve 4, Switzerland}
\affiliation{Theoretical Physics Department, CERN, 1211, Geneva 23, Switzerland}
\emailAdd{matijn.francois@unige.ch, alba.grassi@cern.ch}

\abstract{It is well established that the spectral analysis of canonically quantized four-dimensional Seiberg-Witten curves can be systematically studied via the Nekrasov-Shatashvili functions. In this paper, we explore another aspect of the relation between $\mathcal{N}=2$ supersymmetric gauge theories in four dimensions and operator theory. Specifically, we study an example of an integral operator associated with Painlev\'e equations and whose spectral traces are related to correlation functions of the 2d Ising model. This operator does not correspond to a canonically quantized Seiberg-Witten curve, but its kernel can nevertheless be interpreted as the density matrix of an ideal Fermi gas. Adopting the approach of Tracy and Widom, we provide an explicit expression for its eigenfunctions via an $\o2$ matrix model. We then show that these eigenfunctions are computed by surface defects in $\su2$ super Yang-Mills in the self-dual phase of the $\Omega$-background. Our result also yields a strong coupling expression for such defects which resums the instanton expansion. Even though we focus on one concrete example, we expect these results to hold for a larger class of operators arising in the context of isomonodromic deformation equations.}

\begin{document}

\maketitle

\flushbottom

\section{Introduction}

Building upon the work of Seiberg and Witten \cite{sw1,sw2}, important results have been obtained for $\mathcal{N}=2$ supersymmetric gauge theories in four dimensions.
One remarkable achievement is the exact evaluation of the path integral, made possible thanks to localization techniques and the introduction of the $\Omega$-background  \cite{Moore:1997dj,Losev:1997tp,n,no2}. This led to the discovery of a new class of special functions, the so-called Nekrasov functions \cite{n,no2}, which today have found a wide range of applications in various fields of mathematics and theoretical physics. Examples of such functions are given in \eqref{nek4d} and \eqref{def21}.
Despite the exceptional control Nekrasov functions grant us over the weak gauge coupling regime, a
strong coupling expansion requires alternative methods. This is one of the motivations behind the present work. In addition, this simultaneously explores a particular extension of the correspondence relating $\mathcal{N}=2$ supersymmetric gauge theories in four dimensions to the spectral theory of quantum mechanical operators on the space of square-integrable functions $L^2(\IR)$.

The first concrete example of such correspondence was obtained in \cite{ns}, where the authors showed how to use $\mathcal{N}=2$ supersymmetric gauge theories in the four-dimensional Nekrasov-Shatashvili (NS) phase of the $\Omega$-background $(\epsilon_2=0)$ to solve the spectral theory of a certain class of ordinary differential equations.
For example, the quantization condition for the operator spectrum corresponds to the quantization of the twisted superpotential
\cite{ns,mirmor,mirmor2,Zenkevich2011,Nekrasov:2011bc}, while the eigenfunctions are computed from the surface defect partition functions \cite{Kozlowski:2010tv,Alday:2010vg,Gaiotto:2014ina,Kanno:2011fw,gmn,Sciarappa:2017hds,Jeong:2021rll,Jeong:2023qdr,Jeong:2018qpc,Jeong:2017pai, Bonelli:2022ten,Piatek:2017fyn,Alday:2009fs,Drukker:2009id}. 
More recently, explicit expressions for the Fredholm determinants \cite{ggm,ghn} and the connection coefficients \cite{Bonelli:2022ten} of such differential equations have been derived in closed form.\footnote{A rigorous derivation of some of these results can be found in \cite{Lisovyy:2022flm}. A q-perturbative approach to connection coefficients is also discussed in \cite{Hollands:2021itj,Hollands:2017ahy}, see also \cite{Jeong:2018qpc}.} See also \cite{Gaiotto:2014bza, Ito:2017iba, Yan:2020kkb,Imaizumi:2021cxf,Hollands:2013qza,kpt,Fioravanti:2019vxi, Hollands:2019wbr,Ito:2021sjo} for related development in the context of WKB and four-dimensional gauge theories, and \cite{Grassi:2018bci}
for a different quantization scheme in higher-rank gauge theories. All the operators appearing in the above setup can be obtained via the canonical quantization of four-dimensional Seiberg-Witten (SW) curves \cite{adkmv,mirmor,ns}, or equivalently, by considering the semiclassical limit of BPZ equations \cite{Belavin:1984vu}.

In this paper, we explore another facet of the interplay between spectral theory and supersymmetric gauge theories.  On the gauge theory side, we focus on four-dimensional $\mathcal {N}=2$ gauge theories in the {\it self-dual} phase of the $\Omega$-background ($\epsilon_1=-\epsilon_2=\epsilon$), while on the operator theory side we study a class of operators which do \emph{not} correspond to canonically quantized four-dimensional SW curves.
These operators originally appeared in the framework of isomonodromic deformation equations \cite{wu1,Widom:1996ff,Tracy:1998aa,zamo,Tracy:1995ax}. Their relevance in the context of four-dimensional supersymmetric gauge theories and topological string theory was pointed out in \cite{bgt, bgt2, Bonelli:2017gdk,Gavrylenko:2023ewx}, in close connection with the TS/ST duality \cite{ghm, cgm2, kama, kmz} and the isomonodromy/CFT/gauge theory correspondence \cite{gil1,gil,bsu,Iorgov:2014vla,Bershtein:2018srt,Bershtein:2014yia}. Geometrically, we construct such operators from mirror curves to toric CY manifolds, after implementing a suitable canonical transformation combined with a special scaling limit \cite{bgt, bgt2}, see \autoref{sec:4ddiff} and \autoref{sec:topor}.

Here we focus on a specific example of such an operator which is associated with the Painlev\'e $\rm III_3$ equation, and whose spectral traces compute correlation functions in the 2d Ising model \cite{wu1,zamo,gm}.
Its integral kernel $\rho(x,y)$ on $\IR$ reads\footnote{We refer to \eqref{ointro} as a Fermi gas operator, since it corresponds to the density matrix of an ideal Fermi gas in the external potential $8 t^{1/4} \cosh \left(x\right)$.} 
\be
\label{ointro}
\rho(x,y)={\re^{-4 t^{1/4}\cosh x}\re^{-4 t^{1/4}\cosh y}\over \cosh\left(x-y\over 2\right)} \, ,
\ee
and the corresponding Fredholm determinant $\det\left(1+\kappa\rho\right)$  computes the tau function of Painlev\'e $\rm III_3$ \cite{wu1}.
For $t>0$, the kernel \eqref{ointro} is positive and of trace class on $L^2(\IR)$; hence, the corresponding operator has a discrete, positive spectrum $\left\{E_n\right\}_{n\geqslant 0}$, with  real-valued, square-integrable eigenfunctions $\left\{\varphi_n(x,t)\right\}_{n\geqslant 0}$,
\begin{equation}
\label{eq:IntEqIntro}
    \int_\mathbb{R} \mathrm{d}y \ \rho\left(x, y\right) \varphi_n \left(y, t \right) = E_n \varphi_n\left(x, t \right) \, .
\end{equation}
As we review in \autoref{sec:4ddiff}, the spectrum is computed by the Nekrasov function of 4d, $\mathcal{N}=2$, $\su2$ super Yang-Mills (SYM) in the self-dual phase of the $\Omega$-background \cite{bgt}, see \eqref{etosigma} and \eqref{QC2}.
The purpose of this paper is to study the eigenfunctions of \eqref{ointro} and relate them to surface defects in the same gauge theory.
Specifically, we find that the eigenfunctions $\varphi_n\left(x, t \right)$ are the Zak transform of the sum of two partition functions of surface defects, which are simply related by a change in some parameters.
See equation \eqref{eiged}  for the explicit expression. We test this proposal numerically in \autoref{sec:numeig}.
The relevance of this result is twofold. On the one hand, it provides an efficient description of eigenfunctions of \eqref{ointro} at small $t$.
On the other hand, it offers a tool to resum both the instanton expansion and the $\epsilon$ expansion in the defect partition function, hence allowing to probe the gauge theory at strong coupling (large $t$). 
This is possible because, following \cite{Tracy:1995ax,Marino:2016rsq}, we can provide an alternative matrix model description for such eigenfunctions \eqref{eigintro}, which is exact in $t$ and hence exact in the gauge coupling.

\section{Summary}

\subsection{Results}

\begin{figure}[ht]
    \begin{center}
    \includegraphics[width=0.5\textwidth]{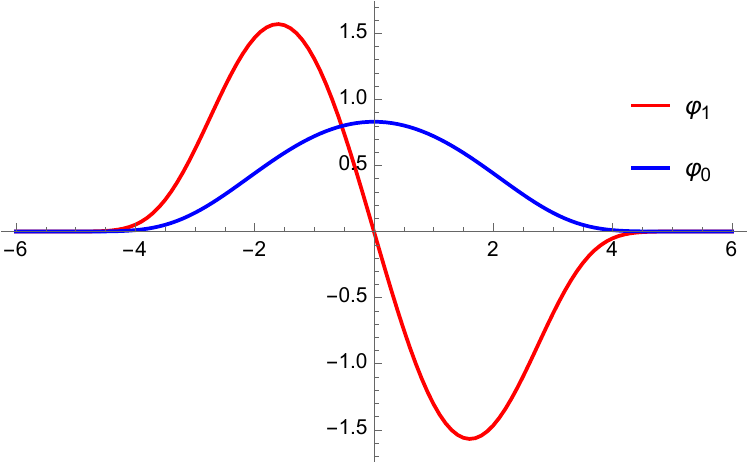}
    \caption{The first (blue) and second (red) eigenfunction of \eqref{ointro} computed using the surface defect expression in \eqref{eiged} for $t = 1 / 100 \pi ^8$.}
    \label{figeig}
    \end{center}
\end{figure}

The paper can be summarized as follows.
Adopting the approach of \cite{Tracy:1995ax}, which is nicely summarized in \cite[sec.~2.3]{Marino:2016rsq}, we construct  eigenfunctions of \eqref{ointro} from 
expectation values of a determinant-like expression,
\be
\label{eigintro}
\Xi_{\pm }\left(x,t,\kappa\right)
=
\re^{-4 t^{1/4}\cosh x} \re^{\pm x/2} \sum_{N\geqslant0}(\pm \kappa)^N  \Psi_N \left( \re^x , t \right) \, ,
\qquad \qquad
x \in \IR \, ,
\ee
\be\label{psiNintro} \ba \Psi_N(z,t)
=
{1\over   N! }\int_{\IR^N_{>0}}\left(\prod_{i=1}^N{\rd z_i  \over z_i} {z-z_i\over z+z_i} ~ \exp\left(-{4 t^{1/4}} \left( z_i + z_i^{-1}\right)\right) \prod_{j = i+1}^N \left({ z_i-z_j \over z_i + z_j }\right)^2
\right) \, . \ea\ee
More precisely, \eqref{eigintro} are  square-integrable eigenfunctions $\varphi_n$ of \eqref{ointro} if we set $\kappa=-E_n^{-1}$, where $E_n$ is an eigenvalue of the operator \eqref{ointro},
\be \varphi_n(x,t)=\Xi_+\left(x,t,-{1\over E_n}\right)=(-1)^n \Xi_-\left(x,t,-{1\over E_n}\right) \, . \ee
In \autoref{sec5}, we show that \eqref{eigintro} and \eqref{psiNintro} are explicitly related to surface defects in 4d, $\mathcal{N}=2$, $\su2$ SYM in the self-dual phase ($\epsilon_1=-\epsilon_2=\epsilon$) of the $\Omega$-background\footnote{This is in line with the generic expectation that matrix model averages are related to partitions functions in the presence of D-branes \cite{adkmv,em,Gukov:2011qp,Maldacena:2004sn,Borot2012,Kozcaz:2010af,Gaiotto:2013sma}.}.
We consider the surface defect which is engineered using the open topological vertex with a D-brane on the external leg, see \autoref{app:vertex} for details. Using the explicit vertex expression of \autoref{app:vertex}, one can see that this corresponds to the special case of a 2d/4d defect called a type II defect in \cite[sec.~2.3.3]{Sciarappa:2017hds}\footnote{Following \cite[sec.~2.3.3]{Sciarappa:2017hds}, the two-dimensional $\mathcal{N}=(2,2)$ theory here consists of 2 free chiral multiplets living on a disc. See also \cite{Bullimore:2014awa, Pan:2016fbl,Gorsky:2017hro}.}.
Hence, we denote its partition function by $Z^{\rm II}_{\rm tot} (q, t, \sigma)$. The explicit expression is given in \eqref{def21} and \eqref{ZIItot}. In the gauge theory, we typically use
\be
\label{eq:GaugeTheoryNotation}
q = {y \over 2 \epsilon} \, ,
\qquad \qquad
t = \left({\Lambda \over \epsilon}\right)^4 \, ,
\qquad \qquad
\sigma= \mathrm{i} {a\over 2 \epsilon} \, ,
\ee
where $y$ is the position of the defect\footnote{The parameter $y$ corresponds to the insertion of the defect on the Riemann sphere $C$ in the 6d, $\mathcal{N} = (2,0)$ theory on $\mathbb{R}^4 \times C$, and it is a position in that sense. From the perspective of the two-dimensional $\mathcal{N} = (2, 2)$ chiral multiplets, it corresponds to the twisted mass for the $U(1)$  flavour symmetry of the chirals, see e.g.~\cite{Gaiotto:2013sma}.}, $\epsilon=\epsilon_1=-\epsilon_2$ is the $\Omega$-background parameter, $a$ is the Coulomb branch parameter, and $\Lambda \sim \re^{-1/g_{\mathrm{YM}}^2}$ is the instanton counting parameter, with $g_{\rm YM}$ the gauge coupling.
The relation between the determinant like expression \eqref{eigintro} and the defect partition function $Z^{\rm II}_{\rm tot}$ is given in \eqref{finale3} and reads
\begin{multline}
\label{finale3intro}
    \Xi_\pm \left( x , t , \frac{\cos(2 \pi \sigma)}{ 2 \pi} \right)
    =
    \\
    { \re^{3\zeta'(-1)}\re^{4 \sqrt{t}}\over 2^{11/12} \pi^{3/2} t^{3/16} }\int_{\IR} \rd q ~ \re^{\ri {2} q x}\sum_{k \in \IZ} \left(Z^{\rm II}_{\rm tot}\left( \pm q, t, \sigma + k \right) + Z^{\rm II}_{\rm tot}\left(\mp q-{\ri\over 2}, t, \sigma + k +{1\over 2}\right)\right) \, .
\end{multline}
The quantization condition for the energy spectrum of \eqref{ointro} was derived in \cite{bgt}, see \eqref{etosigma} and \eqref{QC2}. By evaluating the defect partition function on the right-hand side of \eqref{finale3intro} at the corresponding quantized values of $\sigma = 1 / 2 + \ri \sigma_n$, we obtain the  eigenfunctions $\varphi_n$ of \eqref{ointro},
\begin{multline}
\label{eiged}
    \varphi_n(x, t)
    =
    \\
    { \re^{3\zeta'(-1)}\re^{4 \sqrt{t}}\over 2^{11/12} \pi^{3/2} t^{3/16} }\int_{\IR} \rd q ~ \re^{\ri {2} q x}\sum_{k \in \IZ} \left(Z^{\rm II}_{\rm tot}\left(q , t, k + \frac{1}{2} + \ri \sigma_n  \right) + Z^{\rm II}_{\rm tot}\left(- q - {\ri\over 2}, t, k + \ri \sigma_n \right) \right) ,
\end{multline}
where $\sigma_n \in \mathbb{R}_{>0}$ are solutions to \eqref{QC2}.
The eigenfunctions $\varphi_0$ and $\varphi_1$ are shown in \autoref{figeig}.

In \autoref{sec5} and \autoref{appendix:subsection:finaletofinale}, we show that we can equivalently write  \eqref{finale3intro} as
\begin{multline}
\label{finaleintro}
   \int_{ \IR+\ri \sigma_*}{\rd \sigma}  {\tan \left(2 \pi \sigma\right)\over\left(2 \cos (2 \pi  \sigma )\right)^N} \left(Z^{\rm II}_{\rm tot} \left(q, t, \sigma \right) + Z^{\rm II}_{\rm tot}\left( - q - {\ri \over 2}, t, \sigma+{1\over 2} \right)\right)
    \\
    =  \mathrm{i} {2^{11/12}\sqrt{\pi}{ t^{3/16}  } \over \re^{3\zeta'(-1)}\re^{4 \sqrt{t}} \left( 4 \pi \right)^N} \int_\mathbb{R} {\rd x} ~ \re^{-\ri {2} q x} \re^{ - 4 t^{1/4} \cosh x}  \re^{x/2} \Psi_N \left( \re^{{x}}, t \right) \, ,
\end{multline}
where $\sigma_*$ is such that $0 < \sigma_* < |{\rm Re}(q)|$ if ${\rm Re}(q) \neq 0$, and simply $\sigma_* > 0$ if ${\rm Re}(q)=0$. This choice of $\sigma^*$ guarantees that the integration over $\sigma$ in \eqref{finaleintro} avoids the poles of the integrand. Let us elaborate more on the meaning of \eqref{finaleintro}.
\begin{itemize}
\item[-]The Fourier transform on the right-hand side of \eqref{finaleintro} relates two types of defects  \cite[sect.~2.3.3]{Sciarappa:2017hds}, or more precisely, two phases of the same defect \cite[sect.~4.2]{Jeong:2023qdr}\footnote{Both surface defects are defined by coupling the 4d theory to the same 2d gauged linear sigma model, but in different phases of the  2d theory. The transition between these phases explains the Fourier transform relation between their vacuum expectation values as explained in \cite{Jeong:2023qdr}. We thank Saebyeok Jeong for pointing that out.}.
In particular, while $Z^{\rm II}_{\rm tot}(q, t, \sigma)$ is geometrically engineered in topological string theory by inserting a brane on the external leg of the toric diagram, its Fourier transform with respect to the defect variable $q$ makes contact with a brane in the inner edge of the toric diagram \cite{Jeong:2023qdr}, see also \cite{Aganagic:2001nx,Aganagic:2000gs,adkmv, Kashani-Poor:2006puz}\footnote{We would like to thank A.~Neitzke for useful discussions on this point.}. Following \cite[sect.~2.3.3]{Sciarappa:2017hds}, we refer to the inverse Fourier transform of a type II defect as a type I defect\footnote{Since we consider gauge theories of rank 1 this is equivalent to a type III defect as defined in \cite[sect.~2.3.3]{Sciarappa:2017hds}.}.
Via the AGT correspondence \cite{agt}, the latter is realized in Liouville CFT by considering the five-point function of four primaries with one degenerate field, the so-called $\Phi_{2,1}$ field \cite{Drukker:2009id,Alday:2009fs,Kozcaz:2010af,Dimofte:2010tz,Awata:2010bz, Taki:2010bj}. One can equivalently realize this defect by coupling the four-dimensional theory to a two-dimensional theory, see, for instance, \cite{Jeong:2021rll,Jeong:2023qdr,gmn,Jeong:2018qpc,Kanno:2011fw,Ashok:2017odt,Ashok:2017lko,Sciarappa:2017hds,Pan:2016fbl,Gukov2016} and references therein.

This also means that we could get rid of the inverse Fourier transform on the right-hand side of \eqref{finale3intro} by replacing the partition function of the type II defect $Z^{\rm II}$ with the partition function of the type I defect $Z^{\rm I}$.
The instanton counting-like expression of type I defect can be found, for instance, in \cite{Kanno:2011fw}; however, we will not use such expression here as we will mainly focus on type II defects. 
The reason for this is that, in the latter case, the gauge theoretic expression is represented by a convergent series in  $t$, whose coefficients are exact rational functions of $\sigma$ and $q$. Conversely, for the type I defect, the gauge theoretic expression is more cumbersome, involving double series expansions in both  $t$ and in $q$.

\item[-]The integral over $\sigma$ on the left-hand side of \eqref{finaleintro} is responsible for the change of frame: it brings us from the weakly coupled electric frame, where $Z^{\rm II}_{\rm tot}$ is defined, to the strongly coupled magnetic frame, which is the suitable frame to describe the magnetic monopole point of SYM, see \autoref{TypeIDefect}.
\end{itemize}
In summary, \eqref{finaleintro} means that the matrix model average \eqref{psiNintro} computes the type I surface defect partition function of 4d, $\mathcal{N}=2$, $\su2$ SYM, in the \emph{self-dual phase} of the $\Omega$-background  ($\epsilon_1=-\epsilon_2=\epsilon$), and in the \emph{magnetic frame}. In this identification, $z = \exp\left(x\right)$ is related to the position of the defect and the 't Hooft parameter of the matrix model is identified with the dual period, $N \epsilon = a_D$. 

Note also that \eqref{psiNintro} is exact both in $\Lambda$ and in $\epsilon$; it resums the instanton expansion of the defect partition function and provides an explicit interpolation from the weak to the strong coupling region. The $1/\Lambda$ expansion can be obtained straightforwardly from \eqref{psiNintro} since it corresponds to expanding the matrix model around its Gaussian point, see \cite[sect.~5]{Gavrylenko:2023ewx} and references therein. 
 
\subsection{Derivation}

Let us briefly comment on the derivation of equations \eqref{finale3intro}, \eqref{eiged}, and \eqref{finaleintro}. 
Firstly, we obtained these results by analyzing the large $N$ expansion of the matrix models \eqref{psiNintro} and then extrapolating to finite $N$. Secondly, part of the idea also comes from the open version of the TS/ST correspondence \cite{Marino:2016rsq, Marino:2017gyg}, see \autoref{sec:topor} and \autoref{outlook}. By combining these two approaches we obtained \eqref{finale3intro}-\eqref{finaleintro}, which we further tested numerically. However, we do not have a rigorous mathematical proof of these results.

\hfill

This paper is structured as follows.
In \autoref{review}, we give an overview of the well-established relationship between the modified Mathieu operator and the four-dimensional, $\mathcal{N}=2$, $\su2$ SYM in the NS phase of the $\Omega$-background.
We then present the connection between the operator \eqref{ointro} and the same gauge theory, but in the self-dual phase of the $\Omega$-background.
In \autoref{sec3}, we compute the planar resolvent of \eqref{psiNintro} as well as the planar two-point function and show how the Seiberg-Witten geometry emerges from it. 
In \autoref{TypeIDefect}, we show that the 't Hooft expansion of \eqref{psiNintro} reproduces the $\epsilon$ expansion of the type I self-dual surface defect in the magnetic frame. 
To establish this connection, we rely on two crucial findings. Firstly, according to the results presented in \cite{Kozcaz:2010af}, the $\epsilon$ expansion of the self-dual type I surface defect in the electric frame is determined by topological recursion \cite{eo}. Secondly, the self-dual surface defect (or, more generally, the open topological string partition function) behaves as a wave function under a change of frame \cite{Grassi:2013qva}.  
In \autoref{sec5}, we test \eqref{finaleintro} numerically for finite $N$ and analytically in a $1/N$ expansion, and we verify \eqref{eiged} numerically.

\section{Preparation: spectral theory and 4d, \texorpdfstring{$\mathcal{N}=2$}{N=2} gauge theory}\label{review}

\subsection{Known: differential operators and the NS phase of the \texorpdfstring{$\Omega$}{Omega}-background}\label{oldstory}

Let us start by reviewing the correspondence relating ordinary differential equations to four-dimensional $\mathcal{N} = 2$ gauge theories in the \emph{NS phase} of the $\Omega$-background, where $\epsilon_2 = 0$ and $\epsilon_1 = \epsilon \neq 0$ \cite{ns}. In this paper, we focus on $\su2$ SYM.

 Consider the so-called modified Mathieu operator ${\rm O}_{\rm Ma}$ acting as
\be
\label{mathieu}
{\rm O}_{\rm Ma}~ \varphi(x,t)=\left(-\partial_x^2+\sqrt{t} \cosh \left( x \right) \right)\varphi(x,t)~.
\ee
If $t > 0$,  the operator \eqref{mathieu} has a positive discrete spectrum with square-integrable eigenfunctions. One can make the connection to gauge theory by noting that the modified Mathieu operator corresponds to the canonically quantized SW curve of $\su2$ SYM, if we set $t = \left(\Lambda / \epsilon \right)^4$ as in \eqref{eq:GaugeTheoryNotation}. 
This relationship was exploited in \cite{ns}, showing how gauge theory can be efficiently used to find the spectrum of \eqref{mathieu}. The first ingredient in this relation is the quantization condition of the twisted superpotential which reads
\be
\label{twisteds}
\partial_\sigma F^{\rm NS}(t, \sigma) = \mathrm{i} 2 \pi n \, ,
\qquad \qquad \qquad
n \in \mathbb{N} \, , 
\ee
 where $F^{\rm NS}$ is the NS free energy and $\sigma = \mathrm{i} a / 2 \epsilon$ as in \eqref{eq:GaugeTheoryNotation}. The small $t$ or weak coupling expansion is given by
\begin{multline}
    \label{fnsd}F^{\rm NS}(t, \sigma) = - \psi ^{(-2)}(1-2 \sigma ) - \psi ^{(-2)}(1+2 \sigma) + \sigma^2 \log (t)
    \\
    + \frac{2 t}{ 4\sigma^2 - 1} + \frac{\left( 20 \sigma^2 + 7 \right)t^2}{\left( 4 \sigma^2 - 1 \right)^3 \left( 4 \sigma^2 - 4 \right)}  + \mathcal{O}\left(t^3\right) \, ,
\end{multline}
where $\psi $ is the polygamma function of order $-2$. Higher-order terms in $t$ in \eqref{fnsd} can be computed by using combinatorics of Young diagrams \cite{Nekrasov:2002qd,Bruzzo:2002xf,Flume:2002az}, see \cite{Tachikawa:2013kta}  for a review and further list of references. The resulting expansion in \eqref{fnsd} is convergent\footnote{Beside numerical and physical evidence, we are not aware of a rigorous mathematical proof of this fact, see \cite{ilt,Arnaudo:2022ivo} for related discussions and proofs in other phases of the $\Omega$-background.} when $2\sigma \notin \IZ$. 
Equation \eqref{twisteds} has then a discrete set of solutions $\left\{\sigma_n\right\}_{n\geqslant 0}$ for a fixed value of $t > 0$. The quantum Matone relation \cite{matone,francisco},
\be
\label{quantumm}
E = -t \partial_t F^{\rm NS}\left( t, \sigma \right) \, .
\ee
gives at last the connection to the spectrum $\left\{E_n\right\}_{n\geqslant 0}$ of the modified Mathieu operator \eqref{mathieu}.

There is a parallel development for the eigenfunctions, but one has to consider the four-dimensional partition function with the insertion of a type I defect\footnote{As before, we are following the terminology of \cite[sect.~2.3.3]{Sciarappa:2017hds}. We also note that, since we are considering only rank 1 theories, type I and type III defects as in \cite[sect.~2.3.3]{Sciarappa:2017hds} are identified.} in the NS phase of the $\Omega$-background, see \cite{Kozlowski:2010tv, Alday:2010vg,Kanno:2011fw,Gaiotto:2014ina,gmn,Sciarappa:2017hds,Jeong:2021rll,Jeong:2023qdr} and references there.

\subsection{New: Painlev\'e kernels and the self-dual phase of the \texorpdfstring{$\Omega$}{Omega}-background}\label{sec:4ddiff}

In this work, we consider another class of operators whose spectral properties are encoded in the gauge theory partition functions in the \emph{self-dual phase} of the $\Omega$-background, where $\epsilon_1+\epsilon_2=0$. We focus on the four-dimensional, $\mathcal{N}=2$, $\su2$ SYM. In this case, the relevant operator $\rho$ has the integral kernel $\rho(x,y)$ given by  \cite{bgt}
\be
\label{kernrho}
{\rm \rho}(x,y) = {\sqrt{v(x)}\sqrt{v(y)}\over  \cosh\left({x-y\over 2}\right)} \, ,
\qquad \qquad \qquad
v(x)=\re^{-{8 t^{1/4}} \cosh(\rm x)} \, .
\ee
This operator $\rho$ can be seen as the density matrix of an ideal Fermi gas in an external potential $- \log \left[ v\left(x\right)\right]$ \cite{gm}. We therefore refer to \eqref{kernrho} as a Fermi gas operator. 
Unlike the example of the modified Mathieu operator, the relation of this operator to the quantization of the four-dimensional $\su2$ Seiberg-Witten curves is far from obvious. This connection was originally found in \cite{bgt} as a prediction of the TS/ST correspondence \cite{ghm}. It can be understood as a consequence of the fact that the operator \eqref{kernrho} can be constructed starting from the quantum mirror curve to local $\IF_0$, and implementing a particular scaling limit, as we review in \autoref{sec:topor}. However, we do not yet have a direct method to derive \eqref{kernrho} using any quantization scheme that starts from the four-dimensional SW curve. We have to rely on the TS/ST correspondence. See also the discussion at the end of \autoref{sec:topor}.

For $t>0$ \eqref{kernrho} is a trace class operator on $L^2(\IR)$ with a positive discrete spectrum $\left\{E_n\right\}_{n \geqslant 0}$, 
\begin{equation}
\label{eq:IntEq}
    \int_\mathbb{R} \mathrm{d}y \ \rho\left(x, y\right) \varphi_n \left(y, t \right) = E_n \varphi_n\left(x, t \right) \, .
\end{equation}
It was shown in \cite{bgt} that the spectrum is given by
\be \label{etosigma}E_n^{-1}=-{1\over 2\pi}\cos\left(2 \pi \left({1\over 2}+\ri{\sigma_n}\right)\right) \, , \ee
where $\sigma_n \in \IR_{>0}$ are solutions to
\be\label{QC2} \sum_{k \in \IZ} Z^{\rm Nek}\left(t, {1\over 2} + \ri \sigma_n + k \right)=0 \, . \ee
This result follows from the identity \cite[eqs~3.26,~3.49,~3.55]{bgt} \cite[eq.~2.19]{bgt2}
\be
\label{dete}
\det\left(1+ {\cos(2\pi \sigma)\over 2\pi}\rho\right)=2^{1/12}\re^{3\zeta'(-1)}t^{-1/16}\re^{4\sqrt{t}}\sum_{k \in \IZ} Z^{\rm Nek}(t, \sigma + k ) \, ,
\ee
which was proved using the theory of Painlev\'e equations.
The function $ Z^{\rm Nek}(t, \sigma)$ in \eqref{QC2} and \eqref{dete} is the Nekrasov partition function in the self-dual phase of the $\Omega$-background,
\be \label{nek4d}\ba Z^{\rm Nek}(t, \sigma) ={t^{\sigma^2}  \over  G(1-2\sigma)G(1+2\sigma)} \left( 1+\frac{t}{2 \sigma ^2}+ \frac{\left(8 \sigma ^2+1\right) t^2}{4 \sigma ^2 \left(4 \sigma ^2-1\right)^2}+\mathcal{O}(t^3)\right) \, , \ea\ee
with $G$ the Barnes G-function. Higher-order terms in $t$ are defined systematically by using combinatorics of Young diagrams, see \cite[eqs.~3.4,~3.6]{ilt} for the precise definition. The convergence of this series is shown in \cite{ilt} for any $t>0$ and fixed $2\sigma\notin \IZ$.
Even though $Z^{\rm Nek}(t, \sigma)$ has poles when $2\sigma\in \IZ$, the sum on the right-hand side of \eqref{dete} removes these poles, and the resulting expression is well defined for any complex value of $\sigma$ \cite{ilt, olegunp}.
Let us also note that  the Nekrasov function is often expressed using $\Lambda$, $a$ and $\epsilon$, which are related to $t$ and $\sigma$ via \eqref{eq:GaugeTheoryNotation}
\be
\label{sigmatoa}
t = \left(\frac{\Lambda}{\epsilon}\right)^4 \, ,
\qquad \qquad \qquad
\sigma={\ri}\frac{a}{2  \epsilon } \, . \qquad \ee 

It is useful to write the Fredholm determinant on the left-hand side of \eqref{dete} by using the spectral traces,
\be 
\det\left(1+ \kappa\rho\right)=\sum_{N\geqslant 0} \kappa^N Z(t, N)~,
\ee
\be \label{zpermutation}Z( t, N)= {1\over N!}\sum_{s\in S_N}(-1)^{{\rm sgn }(s)}\int_{\IR}  \rd^N x\prod_{i=1}^N\rho (x_i, x_{s(i)})~,
\ee
where $S_N$ is the permutation group of $N$ elements. 
The Cauchy identity allows us to write \eqref{zpermutation} as \cite{gm}
\be\label{mm}
Z(t, N)
=
{1\over   N! }\int_{\IR^N_{>0}}\left(\prod_{i=1}^N{\rd z_i  \over z_i} \re^{-{4 t^{1/4}} \left(z_i+z_i^{-1}\right)}\prod_{j = i+1}^N \left({ z_i-z_j \over z_i + z_j }\right)^2
\right) \, ,
\ee
which can be analyzed in the regime $t \to + \infty$, since this corresponds to expanding the potential around its Gaussian point.
It was found in \cite{bgt} that the matrix model \eqref{mm} computes the partition function of $\mathcal{N}=2$, $\su2$ SYM in the self-dual phase of the $\Omega$-background and in the magnetic frame\footnote{The magnetic frame can be obtained from the usual electric frame by an S-duality transformation. It allows us to study the behaviour of the gauge theory near the magnetic monopole point. See \autoref{TypeIDefect} and \cite{Tachikawa:2013kta} for more details.}. The relation to the Nekrasov function \eqref{nek4d} can be obtained from \eqref{dete} and reads 
\cite[eq.~2.28]{bgt2} 
\be
\label{Zex}
Z(t, N) = - \mathrm{i} (4 \pi)^N 2^{1/12} \re^{3\zeta'(-1)} t^{-1/16}\re^{4 \sqrt{t}} \int_{ \IR+\ri \sigma_*}{\rd \sigma}  \frac{\tan \left(2 \pi \sigma\right)}{\left(2 \cos (2 \pi  \sigma )\right)^N} Z^{\rm Nek}\left( t, \sigma \right) ,
\ee
where\footnote{\label{footnotess} One can take $\sigma_* = \mathrm{arccosh} (2 \pi ) / 2 \pi $ as in  \cite{bgt2} or any other $\sigma^* > 0$ as long as the integral over $\sigma$ in \eqref{Zex} does not hit the poles of the integrand.} $\sigma_* > 0$. 
Therefore, the matrix model \eqref{mm} provides a resummation of the instanton expansion in the Nekrasov function \eqref{nek4d}, which is an expansion around $t=0$. In this context, we identify the 't Hooft parameter of the matrix model, $N \epsilon$, with the dual or magnetic period in SW theory\footnote{It is possible to analytically continue the results to non-integer values of $N$ \cite{bgt2, cgm}.}
\be
a_D=N \epsilon~.
\ee
The equality \eqref{Zex} was demonstrated in \cite{bgt,bgt2}.
Finally, we emphasise that \eqref{mm} is exact both in the instanton counting parameter $\Lambda$ and in the $\Omega$-background parameter  $\epsilon$. When we expand \eqref{mm} at large $\Lambda$ while keeping $\epsilon$ and $a_D$ fixed, we obtain an analogous expansion to that found when performing a large-time expansion in isomonodromic deformation equations \cite{ilt, Bonelli:2016qwg,Nagoya:2015cja,Nagoya:2018pgp, Lisovyy:2016qig,Gavrylenko:2020gjb,Gavrylenko:2023ewx}. On the matrix model side, this is an expansion around the Gaussian point. Similarly, if we expand at small $\epsilon$ while keeping $\Lambda$ and $a_D$ fixed, we recover the expansion resulting from the holomorphic anomaly algorithm  \cite{bcov,hk06}.

The goal of this work is to extend these results to the eigenfunctions of  \eqref{kernrho}, which on the gauge theory side corresponds to inserting surface defects. 
As a first observation, we note that the kernel \eqref{kernrho} falls in the class of operators studied in \cite{Tracy:1995ax}, and more recently in \cite[sect.~2]{Marino:2016rsq}. In particular, following \cite{Tracy:1995ax,Marino:2016rsq} we can construct eigenfunctions of \eqref{kernrho} using the matrix model \eqref{mm}. Let us define
\be
\label{ximm}
\Xi_{\pm}\left(x,t, \kappa\right)
=
\re^{-{4 t^{1/4} } \cosh( x)} \re^{\pm x/2}\sum_{N\geqslant0}(\pm \kappa)^N  \Psi_N \left( \re^x , t \right) , \qquad \qquad
x\in \IR~,
\ee
\be\label{psi}\ba   \Psi_N(z,t)
={1\over   N! }\int_{\IR^N_{>0}}\left(\prod_{i=1}^N{\rd z_i  \over z_i} {z-z_i\over z+z_i} ~ \exp\left(-{4 t^{1/4}} \left( z_i + z_i^{-1}\right)\right) \prod_{j = i+1}^N \left({ z_i-z_j \over z_i + z_j }\right)^2
\right) . \ea\ee
The $ \Xi_{\pm}(x,t,\kappa)$ in \eqref{ximm} are then square-integrable eigenfunctions $\varphi_n(x, t)$ of \eqref{kernrho} if we evaluate them at $\kappa=-E_n^{-1}$,
\be
\varphi_n(x, t)
=
\Xi_{+}\left(x,t,-{1\over E_n}\right)
=
\left(-1\right)^n\Xi_{-}\left(x,t,-{1\over E_n}\right) .
\ee 
This can be verified by using \cite[eqs.~2.46, 2.59]{Marino:2016rsq} and  $\varphi_n(x, t)=(-1)^n \varphi_n(-x, t)$.
We will argue in the forthcoming sections that the matrix model with insertion $\Psi_N(z,t)$ corresponds to a surface defect in four-dimensional, $\mathcal{N}=2$, $\su2$ SYM in the {\it{self-dual phase}} of the $\Omega$-background and in the {\it{magnetic frame}}.

\subsection{Comment on blowup equations}\label{sec:bup}

It was first pointed out in \cite{huang1606} that the five-dimensional NS and self-dual partition functions are closely connected, which was subsequently demonstrated using Nakajima--Yoshioka blowup equations in \cite{ggu}.  
The interplay between these two phases of the $\Omega$-background was extended to surface defects in four dimensions in \cite{Jeong:2020uxz,Nekrasov:2020qcq}.  Applications in the context of Painlev\'e equations are discussed in \cite{Lencses:2017dgf, Gavrylenko:2020gjb, Jeong:2020uxz, Nekrasov:2020qcq, Bershtein:2021uts, daCunha:2022ewy}. The relevance of blowup equations in the context of resurgence was also recently investigated in \cite{Gu:2023wum}. 

Given such results, it is natural to wonder whether blowup equations can be used to relate the spectrum and eigenfunctions of \eqref{mathieu} and \eqref{kernrho}.
Regarding the spectrum, the blowup formula presented in \cite[eq.~5.7]{Bershtein:2021uts} reveals a one-to-one correspondence between the solutions  $\{\sigma_n\}_{n\geqslant 0}$ of \eqref{twisteds} and the solutions  $\{\sigma_n\}_{n\geqslant 0}$ of \eqref{QC2}. However, 
to obtain the spectrum we further need the quantum Matone relation \eqref{quantumm} on the Mathieu side and the relation \eqref{etosigma} on the Fermi gas side.
These two relations are very different, and therefore, the spectrum of  \eqref{mathieu} and \eqref{kernrho} is related in a highly non-trivial way.
It would be interesting to see if blowup equations in the presence of defects \cite{Jeong:2020uxz,Nekrasov:2020qcq} could be used to establish a map between the eigenfunctions of these two operators. 
 
\subsection{Two limits of quantum mirror curves}\label{sec:topor}

Both operators \eqref{mathieu} and \eqref{kernrho} can be obtained as different limits of the quantized mirror curve corresponding to the toric Calabi-Yau (CY) manifold known as local $\mathbb{F}_0$, which we review here briefly. 

It is well known that four-dimensional $\mathcal{N}=2$ supersymmetric theories can be engineered by using topological string theory on toric CY manifolds \cite{kkv, Iqbal:2003zz, Klemm:1996bj}.
The partition function of refined topological string theory is then identified with the partition function of a five-dimensional $\mathcal{N}=1$ theory on  $\IR^4\times S^1$ \cite{ikv,ta}. If we shrink the $S^1$ circle we get the 4d theory we are interested in, we refer to \cite{Tachikawa:2013kta} for a review and more references. 
For $\mathcal{N}=2$, $\su2$ SYM the relevant setup is topological string theory on local $\IF_0$.
The mirror curve of local $\IF_0$ is
\be
\label{curve}
\re^x+\re^{-x}+ {1\over m_{\IF_0}}\re^{y}+\re^{-y} + \widehat{\kappa} = 0 \, ,
\ee
where $\widehat{\kappa}$ and $m_{\IF_0}$ are the complex moduli. 
The quantization of this curve \cite{adkmv,acdkv} leads to the operator
\be
\label{OF0}
\mathrm{O}_{\IF_0}=\re^\x+\re^{-\x}+ {1\over m_{\IF_0}}\re^{\y}+\re^{-\y} \, ,
\qquad \qquad
[\x,\y]=\ri \hbar \, .
\ee
If $\hbar, m_{\IF_0} > 0$  the inverse operator \be \label{rhof0}\rho_{\IF_0}=\mathrm{O}_{\IF_0}^{-1}\ee is of trace class with a positive discrete spectrum\footnote{It is also possible to take $\hbar$ and $m_{\IF_0}$ imaginary within some range. In this case, one still has the trace class property, but the spectrum may no longer be real \cite{Codesido:2016ixn,gmcomp}.} \cite{ghm,kama,Laptev:2015loa}. Hence, a natural object to consider is its Fredholm determinant
\be
\det \left( 1 + \widehat{\kappa} \rho_{\IF_0} \right) \, .
\ee

The operator \eqref{mathieu} can be obtained from \eqref{OF0} by implementing the usual geometric engineering limit \cite{kkv,Klemm:1996bj} where we scale 
\be
\label{limit1}
m_{\IF_0} = \frac{\beta^4 t}{4} \, ,
\qquad \qquad
\widehat{\kappa} = - \frac{4}{\beta^2 \sqrt{t}} + \frac{2 E}{\sqrt{t}} \, ,
\qquad \qquad
\hbar = \beta \, ,
\ee
and take $ \beta \to 0$. In this limit, we obtain the modified Mathieu operator \eqref{mathieu},
\be
\label{limit1op}
\left(\mathrm{O}_{\IF_0}+\widehat{\kappa}\right)\psi(x)=0~\to~ \left({\rm{O}_{\rm Ma}}+E\right)\psi(x) = 0 \, .
\ee
Likewise, the Fredholm determinant becomes
\be
\det \left( 1 + \widehat{\kappa} \rho_{\IF_0} \right)\to \det \left( 1 + E \, {\rm O}_{\rm Ma}^{-1} \right) ,
\ee
and we have an explicit expression for this determinant via the NS functions \cite[sect.~5]{ggm} 
\be
\det \left(1 + E \, {\rm O}_{\rm Ma}^{-1} \right)= C(t){\sinh\left(\partial_{\sigma}F^{\rm NS}(t, \sigma)\right)\over \sin\left(2 \pi \sigma \right)}~,
\ee
where $C(t)$ is a normalization constant and the relation $\sigma \equiv \sigma(t, E)$ is obtained from \eqref{quantumm}.
 
The Fermi gas operator \eqref{kernrho} on the other hand can be obtained from $\rho_{\IF_0}$ by implementing a rescaled limit \cite{bgt}, 
\begin{equation}
\label{dual4d}
    \log \left( m_{\IF_0} \right) = 4 \pi \mathrm{i} \sigma - \frac{2 \pi}{\beta} \log (\beta ^4 t)  \, ,
    \qquad \qquad
    \widehat{\kappa} = 2 \cos \left( 2 \pi \sigma \right) \, ,
    \qquad \qquad
    \hbar =  \frac{\left(2 \pi \right)^2}{\beta} \, ,
\end{equation}
and $\beta\to 0$. This is called ``the dual 4d limit" in \cite{bgt}. 
The scaling \eqref{dual4d} may seem strange at first sight, but it is a natural limit from the point of view of the TS/ST correspondence \cite{ghm}. In the dual 4d  limit, we have
\be
\label{limitdual4do}\det \left(1+\widehat{\kappa} \rho_{\IF_0} \right)\to \det \left(1+\left({\cos(2\pi \sigma)\over 2 \pi}\right)\rho \right) ,
\ee
where $\rho$ is the operator \eqref{kernrho}. 
The determinant at the right-hand side can also be written as the Zak transform of the self-dual Nekrasov function \eqref{dete}.
  
Let us conclude this section by emphasizing that \eqref{mathieu} has a natural interpretation directly within the four-dimensional theory, independently of the five-dimensional quantum curve. In particular, \eqref{mathieu} is the canonical quantization of the four-dimensional SW curve of $\su2$ SYM, which is related to the semiclassical limit of BPZ equations via the AGT correspondence. 
On the other hand for the Fermi gas operator \eqref{kernrho}, we do not have a parallel interpretation at the moment. It may be possible to relate this operator to some other quantization scheme of the four-dimensional SW curve. Probably a scheme similar to the one used in the context of topological recursion \cite{Iwaki:2015xnr,Iwaki:2019zeq,Marchal:2019nsq,Marchal:2019bia}\footnote{We would like to thank M.~Mari\~no and N.~Orantin for useful discussions on this point.}. 

\section{The Seiberg-Witten geometry from the matrix model}\label{sec3}

In this section, we study the 't Hooft expansion of the matrix model \eqref{psi} and show how the Seiberg-Witten geometry emerges from it.
For this purpose it is useful to parameterise $t = \left( \Lambda / \epsilon \right)^4$ as before in \eqref{sigmatoa}
and to introduce the potential\footnote{The potential of the one-dimensional ideal Fermi gas is $- \log \left( v\left( x \right) \right) = V\left( \mathrm{e}^x \right) / \epsilon$.} $V$ such that
\begin{equation}
\label{vpot}
    v\left(\log \left(z\right)\right) = \exp\left( - \frac{V\left( z \right)}{\epsilon}\right) \, ,
    \qquad \qquad \qquad
    V \left( z \right) = 4 \Lambda \left( z + \frac{1}{z} \right) \, ,
\end{equation}
and we take $\Lambda, \epsilon > 0$ for convenience.
The matrix models \eqref{mm} and \eqref{psi} can then be studied in a 't Hooft limit where
\begin{equation}
\label{thoft}
    N \to + \infty \, ,
    \qquad \qquad
    \epsilon \to 0 \, ,
    \qquad \qquad
    \lambda = N \epsilon  \, ,
\end{equation}
with the defect insertion parameter $z$, the instanton counting parameter $\Lambda$, and the 't Hooft parameter $\lambda$ all kept fixed.

This limit was implemented on the matrix model without insertions \eqref{mm} in \cite{bgt,gm}. In particular, in this limit the eigenvalues of the matrix model distribute along
\be
\left[ \tg, \tg^{-1}\right] \subset \IR_{>0} \, ,
\qquad \qquad \qquad
0 < \tg < 1 \, ,
\ee 
and the 't Hooft parameter $\lambda$ is given by
\be
\label{mmlam}
\lambda=2\Lambda \frac{\mathrm{K}\left(1-\tg^4\right) \left[ 2 \mathrm{E}\left(\tg^4\right) - \left( 1 - \tg^4\right) \mathrm{K}\left(\tg^4\right)\right]-\pi }{ \pi  \tg \mathrm{K}\left(\tg ^4\right)} \, ,
\ee
where $\mathrm{K}$ and $\mathrm{E}$ are the complete elliptic integrals of the first \eqref{eq:EllipticF} and second \eqref{eq:EllipticE} kind, respectively\footnote{See \autoref{app:EllipticIntegrals} for our conventions on elliptic integrals.}.
Later we will use the inversion of this relation for small $\lambda$,
\be\label{gtolsmall} \tg= 1-\frac{\sqrt{\lambda \over \Lambda}}{\sqrt{2}}+\frac{\lambda }{4 \Lambda}-\frac{\lambda ^{3/2}}{32 \sqrt{2}\Lambda ^{3/2}}-\frac{\lambda ^2}{64 \Lambda ^{2}}+ \mathcal{O}\left(\lambda^{5/2}\right) \, . \ee
In the 't Hooft limit \eqref{thoft}, we have the following behaviour
\be\label{zhof} \log Z( t, N)\simeq \sum_{g\geqslant 0}\epsilon^{2g-2} F_g\left( \Lambda, \lambda \right) \, , \ee
where $\simeq$ stands for asymptotic equality. The first two terms read
\be\label{f0mm} {\rd^2 \over \rd \lambda^2} F_0  = -2 \pi{ \mathrm{K}(\tg^{4})\over \mathrm{K}(1-{ \tg^{4}})} \, , \ee
\be \label{F1mm} F_1 =-\frac{1}{4} \log \left( \mathrm{K}\left(1-\frac{1}{\tg^4}\right) \mathrm{K}\left(1-\tg^4\right)\right)-\frac{1}{6} \log \left(\frac{1}{\tg^2}-\tg^2\right)+{\rm constant} \, ,\ee
and higher order terms can also be computed systematically \cite{bgt}.
 
Let us now consider the model with insertions \eqref{psi}. In the 't Hooft limit \eqref{thoft}, we have the following behaviour \cite{Kostovo2,kos39,kos40,kos,ek1,Eynard:1995nv}
 \be \label{t2}\log \left({ {\sqrt{v(z)}} \Psi_N(z,t)\over Z( t, N )} \right)  \simeq \sum_{n \geqslant 0} \epsilon^{n-1} \CT_n \left( z \right) \, . \ee
The leading-order term $\CT_0$ is related to the even part of the planar resolvent\footnote{The planar resolvent is defined and computed explicitly further on in \autoref{sec:planarw0}.} $ \omega_+^0$ \cite[eq.~3.35]{Marino:2016rsq},
\be \label{eq:def:MatrixModelT0} \mathcal{T}_0 \left( z \right) = - 2 \Lambda \left(z + \frac{1}{z}\right)+2 \lambda \int_{\infty}^z \rd z_1  ~\omega_+^0\left( z_1 \right) \, , \ee
and the subleading-order term $ \CT_1$ is given by \cite{Marino:2016rsq, Eynard:1995nv, ek1}
\be
\label{ct1}
\CT_1\left( z \right) = {2} \int^{{z}}_{\infty}\int^{{z}}_{\infty} \rd z_1 \rd z_2 \ W_{++}^0\left( z_1 , z_2 \right) \, ,
\ee
where $W_{++}^0$ is the even part of the planar two-point correlator. It can be expressed explicitly in terms of $\tg$ as \cite{Eynard:1995nv}, \cite[eq.~3.43]{Marino:2016rsq}
\begin{multline}
\label{eq:def:EvenPlanar2PointFun}
    W_{++}^0\left( z_1 , z_2 \right) 
    = 
    \\
    \frac{ \tg^2 + \tg^{-2} - 2 \tg^{-2} \frac{\mathrm{E}\left(1-\tg^4\right)}{\mathrm{K}\left(1-\tg^4\right)} -\left(z_1^2+z_2^2\right) \left(1-\frac{\left(\sqrt{\left(z_1^2-\tg^{-2}\right) \left(z_1^2-\tg^2\right)}-\sqrt{\left(z_2^2-\tg^{-2}\right) \left(z_2^2-\tg^2\right)}\right)^2}{\left(z_1^2-z_2^2\right)^2}\right)}{8 \sqrt{\left(z_1^2-\tg^{-2}\right) \left(z_1^2-\tg^2\right)} \sqrt{\left(z_2^2-\tg^{-2}\right) \left(z_2^2-\tg^2\right)}} \, .
\end{multline}

\subsection{The planar resolvent}
\label{sec:planarw0}

The planar resolvent is
\begin{equation}
    \omega^0(z)
    =
    \lim_{N \to + \infty} {1\over N} \left\langle \sum_{n = 1}^N \left( \frac{1}{z - z_n} \right) \right\rangle \, ,
\end{equation}
where the normalized expectation value is with respect to the matrix model without insertions $Z \left( t, N \right)$ \eqref{mm} and the $z_n$ are the eigenvalues over which one integrates in \eqref{mm}.
At large $z$, one finds
\be
\label{omw}
    \omega^0(z) = \frac{1}{z} + \frac{\langle W \rangle}{z^2} + \mathcal{O}\left( \frac{1}{z^3} \right) \, ,
    \qquad \qquad \qquad
    \langle W\rangle = \lim_{N \to + \infty} \frac{1}{N} \left\langle \sum_{n = 1}^N z_n \right\rangle \, ,
\ee
and we refer to $\langle W\rangle$ as a Wilson loop by analogy with \cite{Suyama:2012uu}. 
It is useful to split the planar resolvent in an even and an odd part,
\be
\omega^0(z)=\omega_+^0(z) +{z}\omega_-^0(z) \, ,
\ee
where $\omega_\pm^0 \left( z \right)$ are both even in $z$.

The even part of the planar resolvent $\omega_+^0$ for the model \eqref{psi} has the following integral form \cite[eq.~4.16] {Eynard:1995nv},
\begin{equation}
\label{eq:def:evenpartplanarresolvent}
    \lambda \omega_+^0(z)
    =
    - \frac{\mathrm{i}}{2} \oint_\mathcal{C} \frac{\mathrm{d} y }{2 \pi \mathrm{i}} \left(\frac{V'(y) y}{z^2 - y^2}\right) \frac{\sqrt{ z^2 - \tg^{-2} }\sqrt{ z^2 - \tg^2 }}{\sqrt{\tg^{-2} - y^2 }\sqrt{ y^2 - \tg^2}} \, ,
\end{equation}
where $\mathcal{C}$ is an anticlockwise contour around the branch cut from $\tg$ to $\tg^{-1}$, which does not include the two poles at $y = \pm z$.
In the matrix model $\Psi_N \left( z , t \right)$ \eqref{psi}, we naturally have $z > 0$. However, it is useful to consider more generally $z \in \mathbb{C}$ from now on, and \eqref{eq:def:evenpartplanarresolvent} makes indeed sense for complex values of $z$ as well \cite{Eynard:1995nv}.

If $z^2 \in \mathbb{C} \setminus [\tg^2, \tg^{-2}]$, we can write write \eqref{eq:def:evenpartplanarresolvent} as
\begin{equation}
\label{eq:evenpartplanarresolventOOBC}
    \lambda \omega_+^0(z)
    =
    \left( \frac{2 \Lambda}{\pi} \right) \sqrt{ z^2 - \tg^{-2} }\sqrt{ z^2 - \tg^2 } \int_\tg^{\tg^{-1}}\mathrm{d} y \, \left(\frac{y - 1 / y}{z^2 - y^2} \right) \frac{1}{\sqrt{\left(\tg^{-2} - y^2 \right) \left( y^2 - \tg^2\right)}} \, .
\end{equation}
where we used the form of the potential given in  \eqref{vpot}. 
The integrand in \eqref{eq:evenpartplanarresolventOOBC} can be decomposed in partial fractions,
\begin{multline}
\label{eq:PartFracDecompRes}
    \int_b^a \mathrm{d} y \left( y - \frac{1}{y} \right) \left(\frac{1}{z^2 - y^2} \right) \frac{1}{\sqrt{\left(a^2 - y^2 \right) \left( y^2 - b^2\right)}} 
    =
    \\
    \frac{1}{2 z^2}\left[ \left( 1 - z^2 \right) (\mathcal{I}(z,a,b) + \mathcal{I}(-z,a,b)) - 2 \, \mathcal{I}(0,a,b) \right] \, ,
\end{multline}
where $0 < b < a$ are any positive real numbers and we defined
\begin{equation}
    \mathcal{I}\left(z,a,b\right) =\int_b^a \mathrm{d} y \left(\frac{1}{y-z}\right) \frac{1}{\sqrt{\left(a^2 - y^2\right) \left( y^2 - b^2\right)}} \, .
\end{equation}
Using \cite[eqs.~256.39,~257.39]{byrd1971handbook}, one finds for $z \notin \left[ b, a \right]$,
\begin{equation}
\label{iden}
    \begin{aligned}
        \mathcal{I}\left(z,a,b\right)
        & = \left(\frac{2}{a+b}\right) \left(\frac{1}{b-z}\right) \int_0^{\mathrm{K(k^2)}} \mathrm{d}v \left( \frac{ 1 - k \, \mathrm{sn}^2\left(v \middle| k^2 \right)}{1- k \left(\frac{z+b}{z-b}\right) \mathrm{sn}^2\left(v \middle| k^2 \right)} \right)
        \\
        & = \left(\frac{2}{a+b}\right) \left(\frac{1}{a-z}\right) \int_0^{\mathrm{K(k^2)}} \mathrm{d}v \left( \frac{1-\left(-k\right) \, \mathrm{sn}^2\left(v \middle| k^2 \right)}{1-k \left(\frac{a+z}{a-z}\right) \mathrm{sn}^2\left(v \middle| k^2 \right)} \right)
    \end{aligned}    
\end{equation}
where $k$ is the elliptic modulus given by
\begin{equation}
    k = \frac{a - b}{a + b} \, ,
\end{equation}
and $\mathrm{sn}\left( v \middle| k^2 \right)$ is the Jacobi elliptic function known as the sine amplitude \eqref{eq:JacobiSN}. From \cite[eq.~340.01]{byrd1971handbook},
\begin{equation}
    \int \mathrm{d}v \left(\frac{1 - \alpha_1^2 \mathrm{sn}^2\left(v \middle| k^2 \right)}{1 - \alpha^2 \mathrm{sn}^2\left(v \middle| k^2 \right)}\right) = \frac{1}{\alpha^2} \left[\left(\alpha^2-\alpha_1^2\right) \Pi \left(\alpha^2, \phi \middle| k^2\right) +\alpha_1^2 v \right] \, ,
    \qquad
    v = \mathrm{F}\left(\phi\middle|k^2\right) \, ,
\end{equation}
where $\mathrm{F}$ and $\Pi$ are the incomplete elliptic integrals of the first \eqref{eq:EllipticF} and third \eqref{eq:EllipticPi} kind, respectively. It is useful to note that $v = 0$ corresponds to $\phi = 0$ and $v = \mathrm{K}\left(k^2\right)$ corresponds to $\phi = \pi / 2$. In the end, this gives
\begin{equation}
\label{iden2}
    \begin{aligned}
        \mathcal{I}\left(z,a,b\right) & = \left(\frac{2}{a+b}\right) \left(\frac{1}{b^2-z^2}\right) \left[ 2 b \Pi \left( \left( \frac{z + b}{z - b} \right) k \middle| k^2 \right) + (z - b) \mathrm{K} \left( k^2 \right) \right]
        \\
        & = \left(\frac{2}{a+b}\right) \left(\frac{1}{a^2-z^2}\right) \left[ 2 a \Pi \left( \left( \frac{a + z}{a - z} \right) k \middle| k^2 \right) + (z - a) \mathrm{K}\left( k^2 \right) \right]
    \end{aligned}
\end{equation}
and hence, we have in \eqref{eq:PartFracDecompRes} for $z^2 \notin \left[ b^2, a^2 \right]$
\begin{equation}
\label{eq:EllipticIntegrals+z&-z}
    \begin{aligned}
         & \mathcal{I} \left(z,a,b\right) +  \mathcal{I}\left(-z,a,b\right)
         \\
         & \qquad = \left(\frac{4 b}{a+b}\right) \left(\frac{1}{b^2-z^2}\right) \left[ \Pi \left( \left( \frac{z + b}{z - b} \right) k \middle| k^2 \right) + \Pi \left( \left( \frac{z - b}{z + b} \right) k \middle| k^2 \right) - \mathrm{K} \left( k^2 \right) \right]
         \\
         &  \qquad = \left(\frac{4 a}{a+b}\right) \left(\frac{1}{a^2-z^2}\right) \left[ \Pi \left( \left( \frac{a + z}{a - z} \right) k \middle| k^2 \right) +  \Pi \left( \left( \frac{a - z}{a + z} \right) k \middle| k^2 \right) - \mathrm{K}\left( k^2 \right) \right]
    \end{aligned}
\end{equation}
as well as
\begin{equation}
\label{eq:EllipticIntegralz=0}
    \begin{split}
        \mathcal{I}\left(0,a,b\right) & = \left(\frac{2}{a+b}\right) \left(\frac{1}{b}\right) \left[ 2 \Pi \left( - k \middle| k^2 \right) - \mathrm{K} \left( k^2 \right) \right]
        \\
        & = \left(\frac{2}{a+b}\right) \left(\frac{1}{a}\right) \left[ 2 \Pi \left( k \middle| k^2 \right) - \mathrm{K} \left( k^2 \right) \right] \, .
    \end{split}
\end{equation}
These particular combinations of elliptic integrals can be reduced to square roots by making use of the following addition formula for $0 < k < 1$ \cite[eq.~117.02]{byrd1971handbook}\footnote{The statement in \cite[eq.~117.02]{byrd1971handbook} is for $\alpha^2 \in \mathbb{R} \setminus \mathcal{H}_{k^2}$, but when $0 < k < 1$, this can be extended to $\alpha^2 \in \mathbb{C} \setminus \mathcal{H}_{k^2}$ by making use of the identity theorem.},
\begin{equation}
\label{eq:AdditionFormulaWholeC}
    \Pi \left( \alpha^2 \middle| k^2 \right) + \Pi \left( \frac{k^2}{\alpha^2} \middle| k^2 \right) - \mathrm{K} \left( k^2 \right) =
    \frac{\pi}{2} \sqrt{\frac{\alpha^2}{\left( 1 - \alpha^2\right)\left(\alpha^2 - k^2\right)}} \, , \qquad  \alpha^2 \in \mathbb{C} \setminus \mathcal{H}_{k^2}~,
\end{equation}
where ${\mathcal{H}_{k^2}}$ is
\begin{equation}
    {\mathcal{H}_{k^2}} =\left[ 0, k^2 \right] \cup \left[ 1, \infty \right] \, .
\end{equation}
Combining \eqref{eq:EllipticIntegralz=0} and \eqref{eq:AdditionFormulaWholeC} gives
\begin{equation}
    \begin{split}
        \mathcal{I}\left(0,a,b\right) & = \left(\frac{2}{a+b}\right) \left(\frac{1}{b}\right)\left(\frac{\pi}{2}\right) \left(\frac{1}{1+k}\right) = \left(\frac{\pi}{2}\right) \left(\frac{1}{a b}\right) \\
        & = \left(\frac{2}{a+b}\right) \left(\frac{1}{a}\right)\left(\frac{\pi}{2}\right) \left(\frac{1}{1-k}\right) = \left(\frac{\pi}{2}\right) \left(\frac{1}{a b}\right) \, ,
    \end{split}
\end{equation}
and the combination of \eqref{eq:EllipticIntegrals+z&-z} and \eqref{eq:AdditionFormulaWholeC} leads to
\begin{equation}
    \begin{split}
        \mathcal{I} & \left(z,a,b\right) +  \mathcal{I}\left(-z,a,b\right)
        \\
        & \qquad \qquad = \left(\frac{4 b}{a+b}\right) \left(\frac{1}{b^2-z^2}\right) \left[ \frac{\pi}{4} \left(\frac{a+b}{b}\right) \sqrt{\frac{b^2-z^2}{a^2-z^2}} \right]  =
            -\frac{\pi}{\sqrt{z^2 - a^2}\sqrt{z^2 - b^2}}
        \\
        & \qquad \qquad = \left(\frac{4 a}{a+b}\right) \left(\frac{1}{a^2-z^2}\right) \left[ \frac{\pi}{4} \left(\frac{a+b}{a}\right) \sqrt{\frac{a^2-z^2}{b^2-z^2}} \right] = -\frac{\pi}{\sqrt{z^2 - a^2}\sqrt{z^2 - b^2}}
    \end{split}
\end{equation}
where $ z^2 \in \mathbb{C} \setminus \left[ b^2, a^2\right]$.

Taking $a^{-1} = b = \tg$ 
and using everything above, we finally find for the even planar resolvent
\begin{equation}
\label{planar4d}
    \boxed{
        \lambda \omega_+^0(z) 
        =
	    \Lambda \left( \left( 1 - \frac{1}{z^2} \right) - \frac{\sqrt{z^2-\tg^{2}}\sqrt{z^2-\tg^{-2}}}{z^2} \right) \, .
    }
\end{equation}
Even though we derive \eqref{planar4d} for $z^2 \in \IC \setminus [\tg^2, \tg^{-2}]$, one can verify  that \eqref{planar4d} holds on the whole complex plane. As a consistency check, we compared the analytical result \eqref{planar4d} against the numerical evaluation of \eqref{eq:def:evenpartplanarresolvent} and found perfect agreement.
One can also see that \eqref{planar4d} has the correct asymptotic behaviour,
\begin{equation}
    \begin{aligned}
        \omega_+^0(z) =
        & \ \mathcal{O}\left(1\right) \, ,
        \qquad \qquad
        && \text{for} \ z \to 0 \, ,
        \\
        \omega_+^0(z) =
        & \ \mathcal{O}\left( z^{-2} \right) \, ,
        \qquad \qquad
        && \text{for} \ z \to \infty \, .
    \end{aligned}
\end{equation}
In addition, from the coefficient of the $z^{-2}$-term in the $\ z \to \infty$ expansion we get a closed-form expression for the Wilson loop 
\eqref{omw},
\be  \label{wilex}\langle W \rangle=\frac{\left( \tg^2-1\right)^2 \Lambda  }{2 \tg^2 \lambda} \, .\ee
Using \eqref{gtolsmall}, we obtain
\be \label{wsmall}\frac{\left(\tg^2-1\right)^2 \Lambda }{2 \tg^2 \lambda}=1 +\frac{\lambda }{16 \Lambda }-\frac{\lambda ^2}{256 \Lambda ^2}+\frac{5 \lambda ^3}{8192 \Lambda ^3}-\frac{33 \lambda ^4}{262144 \Lambda ^4}+\mathcal{O}\left(\lambda ^{5}\right).\ee 
We cross-checked  \eqref{wsmall} 
by expanding the matrix model around its Gaussian point,  similar to what was done in \cite[app.~B]{gm}. 

Using \eqref{planar4d} gives for the leading-order $\mathcal{T}_0$ \eqref{eq:def:MatrixModelT0} of the matrix model \eqref{t2} in the 't Hooft limit \eqref{thoft}
\begin{equation}
\label{fin}
    \partial_z \mathcal{T}_0 \left( z \right) = - 2 \Lambda \left( 1 - \frac{1}{z^2} \right) + 2  \lambda \omega_+^0(z) = - 2\Lambda \frac{\sqrt{z^2-\tg^{2}}\sqrt{z^2-\tg^{-2}}}{z^2}~.
\end{equation}
An important point of \eqref{fin} is that the Seiberg-Witten curve of $\mathcal{N}=2$, $\su2$ SYM,
\be\label{SWc} y^2(z)= - 4 \Lambda ^2 \left(z^2+z^{-2}\right) + {u} \, ,\ee
emerges in the planar limit, provided we identify the following quadratic differentials
\begin{equation}
    \left[ \partial_z \mathcal{T}_0 \left(z\right)  \mathrm{d} z \right]^2 = - \left[ \frac{y \left(z\right)}{z} \mathrm{d} z \right]^2,
\end{equation}
and at the same time relate $\tg$ to $u$ by
\begin{equation}
\label{gu}
    \tg^2+\tg^{-2}={u\over 4 \Lambda^2} \, , 
    \qquad \qquad \qquad
    \tg^{\pm 2} = \left(\frac{u}{8 \Lambda^2}\right) \mp \sqrt{\left(\frac{u}{8 \Lambda^2}\right)^2 - 1} \, .
\end{equation}
In equations \eqref{SWc} and \eqref{gu}, $u$ denotes the vacuum expectation value of the scalar in the vector multiplet of $\mathcal{N}=2$, $\su2$ SYM.

\subsection{The planar two-point function}
\label{subsec:Planar2PointFunction}
 
In the previous section, we showed that the Seiberg-Witten curve \eqref{SWc} naturally emerges when considering the planar resolvent. Here we will see that the Bergman kernel emerges similarly when considering the even part of the planar two-point function. We will see later that this characterises the annulus amplitude in the surface defect. The Bergman kernel is defined as \cite{Awata:2010bz}
\be \label{blrdef}B_{q_1, q_2, q_3}(z_1,z_2)= {1\over 2(z_1-z_2)^2} \left( {2 f(z_1, z_2) + G_{q_1, q_2, q_3}\left(k^2\right) (z_1-z_2)^2\over 2 \sqrt{\sigma (z_1) \sigma(z_2)}}+1\right) \, , \ee
with
\be 
\label{blrdefpart2}
f(z_1, z_2)=\frac{u \left(z_1^2+4 z_2 z_1+z_2^2\right)}{24 \Lambda ^2}+\frac{1}{2} \left(-z_1-z_2\right)-\frac{1}{2} z_1 z_2 \left(z_1+z_2\right) \, , \ee
\be
\label{eq:FrameDepFactor}
G_{q_1, q_2, q_3}\left(k^2\right)
=
\left(q_1-q_3\right) \left[\frac{ \mathrm{E}\left(k^2\right)}{\mathrm{K}\left(k^2\right)}\right] + q_3 - \frac{u}{12\Lambda ^2} \, ,
\qquad \qquad
k^2 = \frac{q_1-q_2}{q_1-q_3} \, ,
\ee
and where $q_i$ are the branch points of $\sigma(z)=-z \left(z^2 - \left(u/ 4 \Lambda^2\right) z + 1 \right)$,
\be q_i \in \left\{0,  \left(\frac{u}{8 \Lambda^2}\right) \mp \sqrt{\left(\frac{u}{8 \Lambda^2}\right)^2 - 1}\right\} = \left\{0, \tg^2, \tg^{-2} \right\} \, .\ee
The choice of the order fixes the choice of frame. What we find is that the relevant order here is 
\be
\label{eq:FrameFactor_MagneticFrame}
q_3 = 0 \, , 
\qquad
q_{2} = \left(\frac{u}{8 \Lambda^2}\right) - \sqrt{\left(\frac{u}{8 \Lambda^2}\right)^2 - 1} = \tg^2 \, ,
\qquad
q_1 = \left(\frac{u}{8 \Lambda^2}\right) + \sqrt{\left(\frac{u}{8 \Lambda^2}\right)^2 - 1}= \tg^{-2} \, .
\ee
As we will discuss later this choice makes contact with the magnetic frame.
One can check that the even part of the planar two-point function \eqref{eq:def:EvenPlanar2PointFun} is related to the Bergman kernel \eqref{blrdef} by
\begin{equation}
     W_{++}^0\left(z_1, z_2\right) = -  z_1 z_2 B_{{q}_1, {q}_2, {q}_3}\left(z_1^2, z_2^2\right) - \frac{1}{4 \left(z_1 +  z_2 \right)^2} \, ,
\end{equation}
Hence, the subleading-order $\mathcal{T}_1$ \eqref{ct1} of the matrix model \eqref{t2} in the 't Hooft limit \eqref{thoft} becomes
\be\label{t1new}\boxed{\mathcal{T}_1 \left(z\right) =2 \int_{\infty}^z \int_{\infty}^z \rd z_1 \rd z_2 \left(-  z_1 z_2 B_{{q}_1, {q}_2, {q}_3}\left(z_1^2, z_2^2\right) - \frac{1}{4 \left(z_1 +  z_2 \right)^2}\right)} \; . \ee

\section{Testing the \texorpdfstring{$\epsilon$}{epsilon} expansion for the type I defect}
\label{TypeIDefect}

From the perspective of the B-model, the partition functions of open and closed topological strings can be defined as objects associated with an algebraic curve, and thus, they depend on a choice of frame, namely a choice of a symplectic basis for the homology of the algebraic curve. The transformation properties of the closed string partition function under a change of frame can be derived from the observation that such a partition function behaves like a wavefunction \cite{Witten:1993ed}. Consequently, the genus $g$ free energies behave as almost modular forms under a change of frame \cite{abk}. The wavefunction behaviour was generalized to the open topological string sector in \cite{Grassi:2013qva}. 

Recall that the partition functions of the four-dimensional gauge theories under consideration are derived from the topological string partition functions via the geometric engineering construction \cite{kkv, Iqbal:2003zz, Klemm:1996bj}. As a result, the same transformation properties hold. 

At the level of terminology, the large radius frame in topological string theory is mapped to the electric frame in the four-dimensional theory. In this frame, the $\mathcal{A}$ cycle and the corresponding A period on the SW curve \eqref{SWc} are chosen to be 
\be
\label{atou}
\Pi_A= {1\over 2 \pi \ri} \oint_{\mathcal{A}} {y(z)\over z} \rd z = {1\over \mathrm{i} \pi }\int_{-\ri \pi}^{\ri \pi} {y(z)\over z} \rd z = \frac{2 \sqrt{8 \Lambda ^2+u}~ {\mathrm E}\left(\frac{16 \Lambda ^2}{8 \Lambda ^2+u}\right)}{\pi} \, ,
\ee
where $y(z)$ is given in \eqref{SWc} and $ \mathrm{E}$ is the complete elliptic integral of second kind \eqref{eq:EllipticE}.
We usually denote $a \equiv \Pi_A$.
Likewise the $\mathcal{B}$ cycle and the corresponding period are
\begin{multline}
     \label{eq:bperiod}  \Pi_B= {1\over 2 \pi \ri} \oint_{\mathcal{B}} {y(z)\over z} \rd z ={1\over \ri  \pi}\int_{\tg}^{\tg^{-1}} {y(z)\over z} \rd z 
     \\
     = { \sqrt{8 \Lambda ^2+u} \over \ri \pi} \left({\mathrm K}\left(\frac{2 u}{8 \Lambda ^2+u}-1\right)-{\mathrm E}\left(\frac{2 u}{8 \Lambda ^2+u}-1\right)\right)
\end{multline}
where $\mathrm{K}$ is the complete elliptic integral of the first kind \eqref{eq:EllipticF}. The $\tg^{\pm 1}$ are roots of the SW curve, $y\left(\tg^{\pm 1}\right) = 0$, and are given in \eqref{gu}.
We usually denote $a_D\equiv \ri \Pi_B$.

On the other hand, the conifold frame in topological strings corresponds to the magnetic frame in the four-dimensional theory. This frame is related to the electric frame by an S-duality which exchanges the $\mathcal{A}$- and $\mathcal{B}$-cycles. For the $\su2$ SYM that we study in this paper, the transformation properties of the partition function under a change of frame were studied in \cite{hk06}. The $\epsilon$ expansion of \eqref{Zex} leads exactly to such transformations, as we discuss below.  

\subsection{The electric frame}

We consider a type I defect in the self-dual phase of the $\Omega$-background ($\epsilon_1=-\epsilon_2=\epsilon$),
and we denote the partition function of this  surface defect by
\be\label{ZDI} Z^{\rm I}(z, t, \sigma)\ee
if we are in the electric frame.
As pointed out in \cite{Kozcaz:2010af}, based on \cite{bkmp, Marino:2006hs}, we can compute these defects via the Eynard-Orantin topological recursion \cite{eo}. 
More precisely, we have 
\be
\label{Zdefe}
\log \left( Z^{\rm I}(z, t, \sigma ) \right) \simeq \sum_{g\geqslant0}\sum_{h\geqslant 1}{\epsilon^{2g-2+h}}\int^z_{\infty} \cdots \int^z_{\infty} W_{g,h}(z_1, \ldots , z_h) \ \rd z_1 \cdots \rd z_h
\ee
where $W_{g,h}(z_1,...z_h)\rd z_1\dots \rd z_h$  is an infinite sequence of meromorphic differentials constructed via topological recursion \cite{eo}, and whose starting point is the underlying SW geometry \eqref{SWc}.  Note that we are implicitly using the dictionary \eqref{sigmatoa} and the SW relation \eqref{atou}
to express $\sigma = \mathrm{i} a / 2 \epsilon$ as a function of the SW parameter $u$. 

For the so-called disk amplitude ${W}_{0,1}$, we have\footnote{Here we take $z > \tg^{-1} > 1$ to simplify the discussion and to make contact with the standard formulas in the literature \cite{Kozcaz:2010af,Awata:2010bz}. Hence we are outside the branch cut on the matrix model side.}
\be
\label{eq:w0}
{W}_{0,1}(z) \ \rd z = - \frac{2 \Lambda}{z}  \sqrt{ \left( z^2 + \frac{1}{z^{2}}\right)-{u\over 4 \Lambda^2}} \ \rd z \, ,
\ee
and we note
\be
\mathcal{W}_{0}^{\rm I}(z, \Lambda, u )=\int^z_{\infty} {W}_{0,1}( z_1 ) \ \rd z_1 \, .
\ee
The annulus amplitude ${W}_{0,2}$ is given by
\be
\label{w1lr}
{W}_{0,2}(z_1, z_2) \ \rd z_1\rd z_2
=
- 2 z_1 z_2  \left(  B_{\widetilde q_1, \widetilde q_2, \widetilde q_3}(z_1^2,z_2^2)- {1\over (z_1^2- z_2^2)^2}\right) \ \rd z_1\rd z_2 \, ,
\ee
where $B_{\widetilde q_1, \widetilde q_2, \widetilde q_3}$ is defined as in \eqref{blrdef}, but the choice of $q_i$'s is different. Here we have
\be
\label{eq:FrameFactor_ElectricFrame}
\widetilde{q_1} = 0 \, , 
\qquad
\widetilde{q_2} = \left(\frac{u}{8 \Lambda^2}\right) - \sqrt{\left(\frac{u}{8 \Lambda^2}\right)^2 - 1} = \tg^2 \, ,
\qquad
\widetilde{q_3} = \left(\frac{u}{8 \Lambda^2}\right) + \sqrt{\left(\frac{u}{8 \Lambda^2}\right)^2 - 1}= \tg^{-2} \, ,
\ee
so that $\widetilde{q_1} = q_3$, $\widetilde{q_2} = q_2$ and $\widetilde{q_3} = q_1$ \eqref{eq:FrameFactor_MagneticFrame}.
We denote 
\be
\mathcal{W}_1^{\rm I}(z, \Lambda, u)= \int^z_{\infty}\int^z_{\infty} {W}_{0,2}(z_1, z_2) \ \rd z_1\rd z_2 \, .
\ee
Hence, to subleading-order \eqref{Zdefe} reads
\be
\label{eq:ZdefeAlt}
\log \left( Z^{\rm I}(z, t, \sigma) \right)
=
{{1\over \epsilon} \mathcal{W}_0^{\rm I}(z, \Lambda, u )+\mathcal{W}_1^{\rm I}(z, \Lambda, u)+\mathcal{O}(\epsilon)} \, .
\ee
Given the spectral curve \eqref{SWc} with $W_{0,1}$ and $W_{0,2} \, $, higher-order terms in the $\epsilon$ expansion \eqref{Zdefe} can be computed recursively by using the topological recursion \cite{eo}. 

\subsection{The magnetic frame}
\label{subsection:DefectInTHeMagneticFrame}

We conjecture that the matrix model \eqref{psi} computes the type I surface defect \eqref{ZDI} in the magnetic frame. 
In this section, we test this proposal in the 't Hooft expansion \eqref{thoft}.

\subsubsection{The partition function without defects}\label{sec:magpf}

It is useful to start by reviewing the 
change of frame in the partition function without defect, which follows from the $\epsilon$ expansion of \eqref{Zex}. 
Using the dictionary \eqref{sigmatoa}, the $\epsilon$ expansion of the Nekrasov function reads
\be
\label{epsilonne}
\log \left( Z^{\rm Nek}( t, \sigma )\right)
\simeq
\sum_{g\geqslant 0}\epsilon^{2g-2}\mathcal{F}_g (\Lambda, a) \, ,
\ee
where $\mathcal{F}_g$ are the genus $g$ free energies of $\su2$ SYM in the electric frame.
Thanks to \eqref{Zex}, we can relate \eqref{epsilonne} to the 't Hooft expansion of the matrix model \eqref{mm},
\be
\label{zhof2}
\log \left( Z(t, N) \right)
\simeq
\sum_{g\geqslant 0}\epsilon^{2g-2} F_g( \Lambda, \lambda) \, .
\ee
It was found in \cite{bgt} that the $F_g$'s in \eqref{zhof2} are the SYM free energies in the magnetic frame.
More precisely,
\be
\label{thooftzz}
\re^{\sum_{g\geqslant 0}\epsilon^{2g-2} F_g(\Lambda, \lambda )}\sim \int_{\ri \IR} {\rd a} \ \re^{- \pi  a  N / \epsilon} \ \re^{\sum_{g\geqslant 0}\epsilon^{2g-2}\mathcal{F}_g (\Lambda, a)} \, , 
\ee
where $\sim$ indicates a proportionality between two (divergent) series\footnote{For sake of notation we omitted the proportionality factor $2^{1/12} \re^{ 3\zeta'(-1)} \left( \Lambda / \epsilon\right)^{-1/4} \re^{4 \left(\Lambda/\epsilon\right)^2}$}. The integral on the right-hand side of \eqref{thooftzz} should be understood as a saddle point expansion. This saddle point expansion characterizes the change of frame in SW theory and topological string \cite{Witten:1993ed}, and it has a direct interpretation from the point of view of modular transformations \cite{abk}. It allows us to make the transition from the weak coupling electric frame, where the Nekrasov function \eqref{nek4d} is defined, to the strong coupling magnetic frame, where the matrix model \eqref{mm} naturally emerges.

By writing the saddle point expansion on the right-hand side of \eqref{thooftzz} explicitly, we get
\be
\label{rhst}
\int_{\ri \IR} {\rd a} \ \re^{- \pi  a  N / \epsilon} \ \re^{\sum_{g\geqslant 0}\epsilon^{2g-2}\mathcal{F}_g (\Lambda, a)}
=
\exp\left[ \frac{1}{\epsilon^2} \left( {\mathcal F}_0 \left( \Lambda, a(\lambda) \right) - {\pi  a(\lambda) \lambda} \right) +\mathcal{O}(1)\right] \, ,
\ee
where $\lambda = N \epsilon$ and $a(\lambda) $ is determined by the saddle point equation 
\be
\label{sadd}
\partial_a {\mathcal F}_0 \left( \Lambda, a \right) = \pi  \lambda \, .
\ee
By using \eqref{nek4d} and the dictionary \eqref{sigmatoa}, we get
\begin{multline}
    \label{ddFnek}\partial_a {\mathcal F}_0( \Lambda, a ) = 2 a (\log (a)-\log (\Lambda )-1)
    \\
    +\frac{4 \Lambda ^4}{a^3}+\frac{30 \Lambda ^8}{a^7}+\frac{480 \Lambda ^{12}}{a^{11}}+\frac{10283 \Lambda ^{16}}{a^{15}}+\frac{1287648 \Lambda ^{20}}{5 a^{19}} + \mathcal{O} \left(\Lambda^{24}\right) \, ,
\end{multline}
and if we make use of \eqref{gu} together with the classical Matone relation,
\be
\label{mat}
u = -\Lambda \partial_{\Lambda} \mathcal{F}_0 \left( \Lambda, a \right)
=
a^2+\frac{8 \Lambda ^4}{a^2}+\frac{40 \Lambda ^8}{a^6}+\frac{576 \Lambda ^{12}}{a^{10}} + \frac{11 752 \Lambda^{16}}{a^{14}} + \frac{286 144 \Lambda^{20}}{a^{18}} +\mathcal{O}\left(\Lambda^{24}\right) \, ,
\ee
we find that $\lambda$ in \eqref{sadd} agrees with \eqref{mmlam} as it should. 
The matching between the two sides of \eqref{thooftzz} was discussed in \cite{bgt}. 
We also note that the classical Matone relation \eqref{mat} can be inverted and one finds the usual expression for the $A$-period of the SW curve \eqref{SWc} given in \eqref{atou}.
Likewise  $\partial_a \mathcal{F}_0$ is identified with the $B$-period of the SW curve\footnote{Hence, we have $a_D = \lambda$.}
\be\label{daF} a_D= {\pi^{-1}} \partial_a \mathcal{F}_0 \left( \Lambda, a \right) ,\ee
where $a_D$ is given in \eqref{eq:bperiod}.

\subsubsection{The partition function with defects}

We are interested in extending the analysis to the 't Hooft expansion \eqref{t2} of the matrix model with insertion $\Psi_N(z,t)$ \eqref{psi}. More precisely, we claim that $\Psi_N(z,t)$ gives the self-dual type I surface defect \eqref{ZDI} in the magnetic frame. As we reviewed above, the change of frame for the partition function is encoded in an integral transform \eqref{thooftzz}. As first shown in \cite{Grassi:2013qva}, this is still the case if one considers the partition function in the presence of surface defects which are engineered via the open topological string partition function, see also \cite{Ashok:2017odt}. 

At the level of the $\epsilon$ expansion,
our conjecture reads
\begin{multline}
\label{thooftdef}
    \exp\left[\sum_{g\geqslant 0}\epsilon^{2g-2} F_g( \Lambda, \lambda ) + \sum_{n \geqslant 0} \epsilon^{n-1} \CT_n(z) \right]
    \\
    \sim   \int_{\ri \IR} {\rd a} \ \re^{- \pi  a N / \epsilon} \ \re^{\sum_{g\geqslant 0}\epsilon^{2g-2}\mathcal{F}_g ( \Lambda, a )}  \re^{\sum_{g\geqslant0}\sum_{h\geqslant 1}{\epsilon^{2g-2+h}}\int^z_\infty \cdots \int^z_\infty W_{g,h}(z_1, \ldots, z_h)\rd z_1 \cdots \rd z_h} \, ,
\end{multline}
where $W_{g,h}(z_1, \ldots ,z_h)\rd z_1 \cdots \rd z_h$ are the electric differentials appearing in the topological recursion setup \eqref{Zdefe}, whereas $\CT_n$ are the  magnetic matrix model coefficients appearing in \eqref{t2},
 \be \label{t2b} \left({ {\sqrt{v(z)}}\Psi_N(z,t)\over Z( t, N )} \right)  \simeq  \exp\left[\sum_{n \geqslant 0} \epsilon^{n-1} \CT_n(z)\right] \, . \ee
Parallel to \eqref{thooftzz}, the integral on the right-hand side of \eqref{thooftdef} should be understood as a saddle point expansion which characterizes the change of frame. Equation \eqref{thooftdef} reads to subleading-order in $\epsilon$
\be
\label{leadingdef}
\epsilon^{-1} \mathcal{T}_0(z)+\mathcal{T}_1(z) + \mathcal{O}\left(\epsilon \right) = \epsilon^{-1} {\mathcal W}_{0}^{\rm I}\left(z, a(\lambda)\right) - \frac{\left[\partial_a{\mathcal W}_{0}^{\rm I}\left(z, a(\lambda)\right)\right]^2}{2  \partial_a^2 \mathcal{F}_{0}\left( a(\lambda)\right)}+{\mathcal W}_{1}^{\rm I}( z, a(\lambda) ) + \mathcal{O}\left(\epsilon \right) \, , 
\ee
where $a$ and $\lambda$ are again related by the saddle point equation \eqref{sadd}. In \eqref{leadingdef}, we already used \eqref{thooftzz} and \eqref{rhst} to get rid of terms involving only the free energies $F_g$ and $\mathcal{F}_g$. We show below that the equality in \eqref{leadingdef} indeed holds order by order in $\epsilon$.

At the leading-order $\epsilon^{-1}$, the matching on the two sides of \eqref{leadingdef} follows directly from \eqref{fin} and \eqref{eq:w0}. 
For the subleading-order $\epsilon^0$, we first note that the Bergman kernel entering in $\mathcal{T}_1$ \eqref{t1new}, can  be written as
\begin{multline}
\label{eq:BergmanKernelChangeOfFrame}
B_{ q_1,  q_2,  q_3}(z_1,z_2) = 
\\
B_{\widetilde q_1, \widetilde q_2, \widetilde q_3}(z_1,z_2)+\frac{\pi \left[{z_1 z_2 \left(\tg^4 z_1-\tg^2 \left(z_1^2+1\right)+z_1\right) \left(\tg^4 z_2-\tg^2 \left(z_2^2+1\right)+z_2\right)} \right]^{-1/2}}{8 \, \mathrm{K}\left(\tg^4\right) \mathrm{K}\left(1-\tg^4\right)}
\end{multline}
where we used \eqref{gu} and \eqref{eq:FrameFactor_ElectricFrame}.
Hence, we can rewrite \eqref{t1new},
\begin{equation}
\label{t1new2}
    \begin{split}
        \mathcal{T}_1 \left( z \right)
        &
        = 2 \int_{\infty}^z \int_{\infty}^z \left(-  z_1 z_2 B_{{q}_1 {q}_2 {q}_3}\left(z_1^2, z_2^2\right) - \frac{1}{4 \left(z_1 +  z_2 \right)^2}\right)\rd z_1 \rd z_2
        \\
        &
        = \mathcal{W}_1^{\rm I} \left( z \right) -\int^z_{\infty}\int^z_{\infty}\left(\frac{\pi  \tg^4}{16 \left(\tg^4-1\right)^2 {\Lambda\mathstrut}^2}\frac{ \partial_{\tg}\partial_{z_1}\mathcal{T}_0(z_1)\partial_{\tg}\partial_{z_2}\mathcal{T}_0(z_2)}{ \mathrm{K}\left(\tg^4\right) \mathrm{K}\left(1-\tg^4\right) }\right)\rd z_1\rd z_2 \, ,
    \end{split}
\end{equation}
which leads then to 
\be\ba   \mathcal{T}_1 \left( z \right) = & \, \mathcal{W}_1^{\rm I} \left( z \right) -\frac{\pi  \tg^4}{16 \left(\tg^4-1 \right)^2 {\Lambda\mathstrut}^2} \frac{\left( \partial_{\tg}\mathcal{T}_0(z)\right)^2}{ \mathrm{K}\left(\tg^4\right) \mathrm{K}\left(1-\tg^4\right) }
\\
= & \, \mathcal{W}_1^{\rm I} \left( z \right) -\left( \partial_{a}\mathcal{T}_0(z)\right)^2\frac{\mathrm{K}\left(\frac{4 \tg^2}{\left(\tg^2+1\right)^2}\right)^2}{ \pi  \left(\tg^2+1\right)^2 \mathrm{K}\left(\tg^4\right) \mathrm{K}\left(1-\tg^4\right)} \, ,
\ea\ee
where we used \eqref{atou}. From \eqref{daF}, we have
\be
\label{ddaF0}
\partial_a^2 \mathcal{F}_0\left(a\right) = \frac{\pi  {\mathrm K}\left(\frac{2 u}{8 \Lambda ^2+u}-1\right)}{{\mathrm K}\left(\frac{16 \Lambda ^2}{8 \Lambda ^2+u}\right)} \, ,
\ee
and by combining \eqref{ddaF0} with the identity 
\be\frac{\mathrm{K}\left(\frac{4 \tg^2}{\left(\tg^2+1\right)^2}\right)}{   \left(\tg^2+1\right)^2 \mathrm{K}\left(\tg^4\right) \mathrm{K}\left(1-\tg^4\right)}=\frac{1}{2   \mathrm{K}\left(\frac{\left(\tg^2-1\right)^2}{\left(\tg^2+1\right)^2}\right)} \ee
we find
\be
\label{eq:TypeIvsMM_MagneticFrame_NLO}
\boxed{\mathcal{T}_1 \left( z \right) = \mathcal{W}_1^{\rm I} \left( z \right) - { \left( \partial_{a}\mathcal{T}_0(z)\right)^2 \over 2 \partial_a^2 \mathcal{F}_0}} \, , 
\ee
which is precisely what we wanted to prove.

To summarize, we have tested \eqref{thooftdef} at leading and subleading-order\footnote{For the part involving only the free energies the all order results follows from \cite{bgt,bgt2}.} in $\epsilon$. The matching of higher orders can be inferred from the application of topological recursion. On the canonical defect side, the fact that higher orders in \eqref{ZDI} satisfy the topological recursion was conjectured in \cite{Kozcaz:2010af}, based on \cite{bkmp, Marino:2006hs} which was recently demonstrated in \cite{Eynard:2012nj}. On the matrix model side instead, the inclusion of topological recursion in our matrix model can be derived from \cite{eo,Borot:2009ia,Borot:2013lpa}. Our computation above shows that the initial data for such recursion are the same on both sides; therefore, matching at all orders is also expected. 

\section{Matrix models, eigenfunctions and the type II defect}\label{sec5}

In this section, we consider the Fourier transform of the matrix model with insertion $\Psi_N\left(\mathrm{e}^x, t\right)$ \eqref{psi}. The corresponding defect in four-dimensional, $\mathcal{N} = 2$, $\su2$ SYM can be geometrically engineered using the open topological string partition function of local $\IF_0$, where we insert a  D-brane on the external leg, see \autoref{app:vertex}. The partition function of the resulting type II defect in the self-dual phase of the $\Omega$-background is
\begin{equation}
\label{def21}
    \begin{split}
        Z^{\rm II}(q, t, \sigma )
        & =
        \exp \left( \frac{\mathrm{i}}{2} q \log \left( t \right) \right) \Gamma \left( - \mathrm{i} q-\sigma +\frac{1}{2}\right) \Gamma \left( - \mathrm{i} q + \sigma + \frac{1}{2}\right) Z_{\rm inst}^{\rm II}(q, t, \sigma) \, ,
        \\
        Z_{\rm inst}^\text{II}\left( q, t, \sigma \right)
        & =
        1 - \left[ \frac{\widetilde{q}}{2 \sigma^2 \left(\widetilde{q}^2 - \sigma^2 \right)}\right] t
        \\
        &
        + \left[ \frac{\widetilde{q} \left( \widetilde{q} + 1 \right)^2 - \widetilde{q} \left( 10 \widetilde{q}^2 + 19 \widetilde{q} + 10 \right) \sigma^2 + \left(  8 \widetilde{q}^2 + 30 \widetilde{q} + 9 \right) \sigma ^4}{4 \sigma^4 \left( 4 \sigma^2 - 1 \right)^2 \left(\widetilde{q}^2 - \sigma^2\right) \left( \left( \widetilde{q} + 1 \right)^2 - \sigma^2\right)} \right] t^2 
   +\mathcal{O}\left(t^3\right) \, ,
    \end{split}
\end{equation}
where we defined for the sake of readability $\widetilde{q} = \mathrm{i} q + 1 / 2$. The variables $q, t, \sigma$ can be expressed in terms of $y, \Lambda, a$ and $\epsilon$ as in \eqref{eq:GaugeTheoryNotation}.
The relation between $Z^{\rm II}$ and the matrix model \eqref{psi} reads\footnote{This also suggests that the defect partition function $Z^{\rm II}\left(q, t,\sigma\right)$ for this theory has an infinite radius of convergence in $t$ for fixed $\sigma$ and $q$ which are away from the poles. For the self-dual partition function without defect, this was rigorously proved in \cite{ilt}, see also \cite{Arnaudo:2022ivo} for more recent developments.}
\begin{equation}
\label{finale}
    \begin{split}
        - \mathrm{i} \, 2^{1/12} \re^{3\zeta'(-1)}t^{-1/16}\re^{4 \sqrt{t}} & \int_{ \IR+\ri \sigma_*}{\rd \sigma} {\tan \left(2 \pi \sigma\right)\over\left(2 \cos (2 \pi  \sigma )\right)^N} Z^{\rm Nek}(t, \sigma) \left(\sum_{s}Z^{\rm II}_{s}(q, t, \sigma)\right)
        \\
        & =  {2\sqrt{\pi}{ t^{1/8}  } \over (4 \pi)^N}\int_\mathbb{R} {\rd x} \, \re^{-\ri {2} q x} \re^{-4 t^{1/4} \cosh \left( x \right) + \frac{x}{2}} \Psi_N(\re^{{x}}, t) \, ,
    \end{split}
\end{equation}
where $ \sigma_* $ is chosen such that  $0 < \sigma_* < |{\rm Re}(q)|$ if $\mathrm{Re}\left( q \right) \neq 0$, and simply $\sigma_* > 0$ if $\mathrm{Re}\left( q \right) = 0$. This guarantees that the integral on the left-hand side does not hit the poles of the integrand.
The sum over $s$  can be seen as a sum over saddle points of the integral over $x$. We find that
\be \label{sigamsum}\left(\sum_{s}Z^{\rm II}_{s}(q, t, \sigma)\right)=Z^{\rm II}\left(q, t, \sigma \right)+ \left(-1\right)^N Z^{\rm II}\left(-q-{\ri\over 2}, t, \sigma \right) \, .\ee
It is convenient to introduce the total partition function as
\be \label{ZIItot} Z^{\rm II}_{\rm tot}( q, t, \sigma )= Z^{\rm Nek}(t, \sigma)Z^{\rm II}(q, t, \sigma ) \, , \ee
 so that \eqref{finale} can be written in a compact form as
\be\label{finalecompact} \boxed{\ba & \int_{ \IR+\ri \sigma_*}{\rd \sigma}  {\tan \left(2 \pi \sigma\right)\over\left(2 \cos (2 \pi  \sigma )\right)^N} \left(Z^{\rm II}_{\rm tot}(q, t, \sigma) + (-1)^N Z^{\rm II}_{\rm tot}\left( -q-{\ri \over 2}, t, \sigma\right)\right)\\
& =  \ri {2^{11/12}\sqrt{\pi}{ t^{3/16}  }\over \re^{3\zeta'(-1)}\re^{4 \sqrt{t}}(4\pi)^N}\int_\mathbb{R} {\rd x} \,
 \re^{-\ri {2} q x} \re^{- 4 t^{1/4} \cosh x+ {x\over 2}}
\Psi_N(\re^{{x}}, t) \, . \ea} \ee
This can equivalently be written as
\be\label{finale2}\ba &\sum_{k \in \IZ} \left(Z^{\rm II}_{\rm tot}(q, t, \sigma + k )+Z^{\rm II}_{\rm tot}\left( -q-{\ri\over 2}, t, \sigma + k + {1\over 2} \right)\right)\\
&=  { 2^{11/12}\sqrt{\pi}{ t^{3/16}  }\over \re^{ 3\zeta'(-1)}\re^{4 \sqrt{t}} }\int_\mathbb{R} {\rd x} \,
 \re^{-\ri {2} q x} \re^{- 4 t^{1/4} \cosh x+ {x\over 2}}
\left[\sum_{N\geqslant 0} \Psi_N(\re^{{x}}, t) \left(\frac{\cos(2\pi\sigma)}{2\pi}\right)^N \right] \, .\ea \ee
One can use an argument based on the Fourier series to get from \eqref{finalecompact} to \eqref{finale2}, while the other direction uses Cauchy's residue theorem. See \autoref{appendix:subsection:finaletofinale} for details. By inverting the Fourier transform \eqref{finale2}, we  have
 \be\label{finale3}\boxed{\ba &\int_{\IR} \rd q \ \re^{\ri {2} q x}\sum_{k \in \IZ} \left(Z^{\rm II}_{\rm tot}\left( q, t, \sigma + k \right) +Z^{\rm II}_{\rm tot}\left( - q - {\ri\over 2} , t, \sigma + k + {1\over 2} \right) \right) \\
 & =  { 2^{11/12} \pi^{3/2} t^{3/16} \over \re^{3\zeta'(-1)}\re^{4 \sqrt{t}} }
 \re^{ -4 t^{1/4} \cosh x+ {x\over 2}}
\sum_{N\geqslant 0} \Psi_N(\re^{{x}}, t) \left({\cos(2\pi\sigma)\over 2\pi}\right)^N  \ea }\ee
Following \autoref{sec:4ddiff}, we get the square-integrable eigenfunctions of \eqref{kernrho} when we evaluate \eqref{finale3} at the values of $\sigma$ which satisfy the quantization condition \eqref{QC2}. That is,
\be\label{eigenfff}\boxed{ \ba \varphi_n(x,t)=& {  \re^{3\zeta'(-1)}\re^{4 \sqrt{t}} \over  2^{11/12} \pi^{3/2} t^{3/16}}\\
& \int_{\IR} \rd q \, \re^{\ri {2} q x}\sum_{k \in \IZ} \left(Z^{\rm II}_{\rm tot}\left( q, t, k + {1\over 2} + \ri \sigma_n \right) + Z^{\rm II}_{\rm tot}\left( - q - {\ri\over 2} , t, k + \ri \sigma_n \right)\right) \ea}\ee
where $\sigma_n$ are solutions of \eqref{QC2}.
In \autoref{figeig}, we plot the right-hand side of \eqref{eigenfff} for the two smallest values of $\sigma_n$ that satisfy the quantization condition \eqref{QC2}.
As a cross-check, we also verified this result by a purely numerical analysis of the operator \eqref{kernrho}, see \autoref{sec:numeig}.

Let us make a few comments on the analytic properties of the gauge theoretic functions. 
\begin{itemize}
\item[-] The  function $Z^{\rm Nek}( t, \sigma )$ has poles when $2\sigma\in \IZ$ and $Z^{\rm II}_{\rm tot}(q, t, \sigma)$ has additional poles when $q$ and $\sigma$ satisfy $ q = \frac{\mathrm{i}}{2} \pm \ri \sigma + \ri \ell$ with $ \ell \in \mathbb{Z}$.
\item[-]  If we are strictly interested only in the spectral problem associated with the integral kernel \eqref{kernrho}, then $q \in \IR$ and $\sigma\in \frac{1}{2} + {\ri} \IR_{>0}$. So these poles are not realized. 
\item[-]  However, we can go beyond this special domain. For example, if we consider the Zak transform of $Z^{\rm Nek}(t, \sigma)$ appearing on the left-hand side of \eqref{dete}, then this has no longer poles in $\sigma$: the summation over $k$ in \eqref{dete}  removes the poles. Likewise, it seems that the summation over integers and the particular combination of defect partition functions appearing in the integrand on the left-hand side of \eqref{finale3} has also the effect of removing the poles.
\end{itemize}

In the forthcoming subsections, we test \eqref{finalecompact} and \eqref{eigenfff} in several ways.

\subsection{Testing \texorpdfstring{$N=0$}{N=0}}

As a first check of \eqref{finalecompact}, we test the $N=0$ case. From \eqref{psi}, one can see that $\Psi_{0}(\re^{{x}}, t) = 1$ so that the right-hand side of \eqref{finalecompact} is
\be
\label{integn}
\left( \ri \frac{2^{11/12}\sqrt{\pi} t^{3/16}}{\re^{3\zeta'(-1)}\re^{4 \sqrt{t}}} \right) \int_{\mathbb{R}} {\rd x} \ \re^{-\ri {2} q x} \re^{-4 t^{1/4} \cosh x + {x\over 2}} = \left( \ri \frac{2^{11/12}\sqrt{\pi} t^{1/16}}{\re^{3\zeta'(-1)}\re^{4 \sqrt{t}}} \right) 2 t^{1/8} K_{\mathrm{i} 2 q - \frac{1}{2}}\left(4 t^{1/4}\right) \, ,
\ee
where $K$ is the modified Bessel function of the second kind.
By expanding at small $t$, we find that the Bessel function has the following structure, 
\be
\label{lhs}
   2 t^{1/8} K_{ \mathrm{i} 2 q-\frac{1}{2}}\left(4 t^{1/4}\right) 
    = F(q,t) + F\left(-q-{\ri\over 2},t\right) \, ,
\ee
for some function $F(q,t)$. For instance, we have when $ \ri 2 q - \frac{1}{2} \notin \IZ $
\be \label{fseri} F(q,t)= \re^{\ri \frac{q}{2} \log (t)} \left[ 2^{ \mathrm{i} 2 q - \frac{1}{2}} \Gamma \left( - \mathrm{i} 2  q + \frac{1}{2}\right)-2^{ \mathrm{i} 2 q + \frac{3}{2}} \Gamma \left( - \mathrm{i} 2 q -\frac{1}{2}\right) \sqrt{t} +\mathcal{O}\left(t\right)\right] \, , \ee
Hence, we already see the structure of the left-hand side of \eqref{finalecompact} appearing.
On the gauge theory side, we can perform the integral at small $t$ by using Cauchy's residue theorem, 
\be \label{gaugeint} \ba & \int_{ \IR+\ri \sigma_*}{\rd \sigma}  \tan \left(2 \pi \sigma\right) Z_\mathrm{tot}^{\rm II}(q, t, \sigma) \\
& = \int_{ \IR+\ri \sigma_*}{\rd \sigma}  \tan (2 \pi  \sigma ) \re^{\ri \frac{q}{2} \log (t)} t^{\sigma ^2}\frac{\Gamma \left(- \ri q-\sigma +\frac{1}{2}\right) \Gamma \left(- \ri q+\sigma +\frac{1}{2}\right)}{G(1-2 \sigma ) G( 1 + 2 \sigma )}\left(1+\mathcal{O}(t)\right)\\
& = \left( \ri \frac{2^{11/12} \sqrt{\pi } t^{1/16}}{\re^{3 \zeta '(-1)} \re^{4 \sqrt{t}}} \right) \re^{\ri \frac{q}{2} \log (t)} \left[ 2^{ \mathrm{i} 2 q - \frac{1}{2}} \Gamma \left( - \mathrm{i} 2 q + \frac{1}{2}\right)-2^{ \mathrm{i} 2 q + \frac{3}{2}} \Gamma \left(- \mathrm{i} 2 q-\frac{1}{2}\right) \sqrt{t} \right.\\
&\qquad\qquad\qquad\qquad\qquad\qquad \qquad \left.+ 2^{ \mathrm{i} 2 q + \frac{5}{2}} \Gamma \left(- \mathrm{i} 2 q - \frac{3}{2}\right) \, t + \mathcal{O}\left(t^{3/2}\right)\right] \, .
\ea\ee
To get the last line in \eqref{gaugeint}, we have included the first instanton correction in $Z_\mathrm{tot}^{\rm II}$ \eqref{ZIItot}, and higher instanton corrections can be treated similarly. The poles contributing to the integral in \eqref{gaugeint} are
\be \sigma = { \ell \over 2} \qquad \text{and} \qquad \sigma={1\over 4} + { \ell \over 2} \qquad \text{with} \qquad \ell \in \IZ~.\ee
By employing the series expansions \eqref{fseri} and \eqref{gaugeint}, we can systematically verify \eqref{finalecompact} for $N=0$, order by order in $t$. 

\subsection{Testing \texorpdfstring{$N=1$}{N=1}}

\begin{table}[t] 
\centering\small
   \begin{tabular}{l l l}
  \\
 $n^{\text{inst}}$ &  \\
\hline
  0 &\underline{2.83}72834788 + \underline{4.7}204648771 i\\
 1 & \underline{2.831}2289304 + \underline{4.7137}559136 i \\
 2 & \underline{2.831322}7416 + \underline{4.713743}4937 i\\
 3 & \underline{2.8313226948} + \underline{4.7137435320} i\\ 
 \hline
    $ I_1 $ &  2.8313226948 + 4.7137435320 i  \\
\end{tabular}
\quad
   \begin{tabular}{l l l}
  \\
 $n^{\text{inst}}$ &   \\
\hline
  0 &\underline{0.0502}35280369 + \underline{0.0181}41366757 i\\
 1 & \underline{0.05024}2014312 + \underline{0.0181116}11547 i \\
 2 & \underline{0.050241915}710 + \underline{0.018111648}299 i\\
 3 & \underline{0.050241915600} + \underline{0.018111648316} i\\ 
 \hline
$ I_1 $ & 0.050241915600 + 0.018111648316  i  \\
\end{tabular}
\quad
   \begin{tabular}{l l l}
  \\
 $n^{\text{inst}}$ &   \\
\hline
  0 &\underline{0.1587}865901507390 i\\
 1 &  \underline{0.15876801}11233355 i \\
 2 &  \underline{0.1587680126408}577 i\\
 3 &  \underline{0.1587680126408951} i\\ 
 \hline
$ I_1 $ &  0.1587680126408951 i  \\
\end{tabular}
\caption{Comparison between the two sides of \eqref{finalecompact} for $N=1$, $t= 1 / 55 \pi^4$ with $q= 1/9 + \mathrm{i} 2 / \sqrt{3}$ (upper left),  $q= 1/ \pi $ (upper right), and $q= \mathrm{i} / 3$ (lower). $I_1$ is the integral \eqref{eqintn1.2} appearing on the right-hand side of \eqref{finalecompact}; $n^{\text{inst}}$ refers to the number of instantons we include in the defect partition function appearing on the left-hand side of \eqref{finalecompact}.
}
\label{tb1}
\end{table}
As a second consistency check of \eqref{finalecompact}, we test the $N=1$ case.
First, we note that by a change of variables, we can rewrite the double integral appearing on the right-hand side of \eqref{finalecompact} as a one-dimensional integral. Let us define
\be \ba & I_1(q,t) = \left( \ri {2^{11/12} \sqrt{\pi } {t}^{3/16}\over \re^{3 \zeta'(-1)} \re^{4 \sqrt{{t}}} \left (  4 \pi \right)} \right) \int_{\mathbb{R}}  {\rd x} \
 \re^{-\ri {2} q x} \re^{- 4 t^{1/4} \cosh x + \frac{x}{2}}
\Psi_1(\re^{{x}}, t).\ea\ee
After some algebra, we get
\be \label{eqintn1.2}\ba & I_1(q,t) = \left( \ri \frac{ 2^{11/12} t^{3/16}}{\sqrt{\pi } \re^{3 \zeta'(-1)} \re^{4 \sqrt{t}}} \right) \times \\
&  \int_{3}^{+\infty} \mathrm{d}U \frac{ \left( \left(\frac{U^2+\sqrt{U^4-10 U^2+9}-3}{2 U}\right)^{- \mathrm{i} 2 q + \frac{1}{2}}- \left(\frac{U^2+\sqrt{U^4-10 U^2+9}-3}{2 U}\right)^{ \mathrm{i} 2 q - \frac{1}{2}}\right) K_{\mathrm{i} 2 q - \frac{1}{2}}\left(4 \sqrt[4]{t} U\right)}{U-U^{-1}} \, .
\ea\ee
One useful observation is that the above integral vanishes when $q = -\ri / 4$, which is in perfect agreement with the left-hand side of \eqref{finalecompact}. Unfortunately, we cannot compute the integral \eqref{eqintn1.2} analytically. Hence, for $N=1$, the test of \eqref{finalecompact} is done numerically and we find perfect agreement. One such test is given in \autoref{tb1}.

\subsection{Testing large \texorpdfstring{$N$}{N} with a 't Hooft limit}
\label{subsection:NToInfty}

Another analytical test of the identity \eqref{finalecompact} consists of comparing both sides in the 't Hooft limit, where just as in \eqref{thoft}
\begin{equation}
\label{eq:THooftScalingNEpsilon}
    N \to + \infty \, , \qquad \qquad \epsilon \to 0 \, , \qquad \qquad \lambda = N \epsilon > 0 \, ,
\end{equation}
and with the 't Hooft coupling $\lambda$ fixed. We will need to use that $q$ and $t$ scale as in \eqref{eq:GaugeTheoryNotation},
\begin{equation}
\label{eq:THooftScalingTQ}
   q = \frac{y}{2 \epsilon} \, ,
   \qquad \qquad \qquad
   t = \left(\frac{\Lambda}{\epsilon}\right)^4 \, , 
\end{equation}
with both the position of the defect $y \in \mathbb{C}$ and the instanton counting parameter $\Lambda > 0$ kept fixed.

The computation of the 't Hooft limit of \eqref{finalecompact} is simplified by using the corresponding statement for the theory without defects, which is given in \eqref{Zex} and was obtained in \cite{bgt, bgt2}. In particular, one can divide both sides of \eqref{finalecompact} by \eqref{Zex} to get
\begin{multline}
\label{eq:FinalePreTHooft} 
    \frac{\int_{\mathbb{R}+\mathrm{i}\sigma_*} \mathrm{d}\sigma \frac{\tan\left(2 \pi \sigma \right)}{\left[ 2 \cos \left( 2 \pi \sigma \right)\right]^N} Z^\text{Nek}\left( \left(\frac{\Lambda}{\epsilon}\right)^4, \sigma \right) \left[ Z^\mathrm{II}\left( \frac{y}{2 \epsilon}, \left( \frac{\Lambda}{\epsilon}\right)^4, \sigma \right) + \left(-1\right)^N Z^\mathrm{II}\left( \frac{- y - \mathrm{i} \epsilon}{2 \epsilon}, \left(\frac{\Lambda}{\epsilon}\right)^4, \sigma \right) \right]}{ \left(2 \pi \right) \int_{\mathbb{R}+\mathrm{i}\sigma_*} \mathrm{d}\sigma \frac{\tan\left(2 \pi \sigma \right)}{\left[ 2 \cos \left( 2 \pi \sigma \right)\right]^N} Z^\text{Nek}\left( \left(\frac{\Lambda}{\epsilon}\right)^4, \sigma \right)}
    \\
    =
    \frac{\sqrt{ \Lambda}}{\sqrt{  \pi \epsilon}} \int_\mathbb{R} \mathrm{d}x \, \exp\left(- \frac{\mathrm{i}}{\epsilon} x y \right) \frac{\exp\left(- 4 \left(\frac{\Lambda}{\epsilon}\right) \cosh\left(x\right) + \frac{x}{2}\right)\Psi_N \left( \mathrm{e}^x , \left(\frac{\Lambda}{\epsilon}\right)^4 \right)}{Z\left( \left(\frac{\Lambda}{\epsilon}\right)^4, N \right)} \, .
\end{multline}
Note that \eqref{eq:FinalePreTHooft} is by \eqref{Zex} equivalent to \eqref{finalecompact}, but rewritten in a form suitable and convenient for the 't Hooft limit \eqref{eq:THooftScalingNEpsilon}.

\subsubsection{The 't Hooft limit on the gauge theory side}

The general pattern of the 't Hooft expansion of the left-hand side in \eqref{eq:FinalePreTHooft} is the same as in \autoref{subsection:DefectInTHeMagneticFrame}. Using that the integration variable $\sigma$ can be related to the Coulomb branch parameter $a$ by \eqref{sigmatoa},
\begin{equation}
    \sigma = \mathrm{i} \frac{a}{2 \epsilon} \, ,
\end{equation}
one expands the logarithm of the Nekrasov partition function $Z^\mathrm{Nek}$ in even powers of $\epsilon$ with the leading-order being $\epsilon^{-2}$. On the other hand, the expansion of the logarithm of the defect partition function $Z^\mathrm{II}$ contains all integer powers of $\epsilon$ starting from $\epsilon^{-1}$,
\be
\label{eq:TypeIIDefectEpsilonExp}
(2\pi)^{-1}Z^{\rm II}\left( \frac{y}{2 \epsilon}, \left( \frac{\Lambda}{\epsilon}\right)^4, \mathrm{i} \frac{a}{2 \epsilon} \right) \simeq \exp \left( \sum_{n\geqslant 0} \mathcal{W}^{\rm II}_n(y) \epsilon^{n-1} \right) \, .
\ee
Hence, the saddles of both integrals on the left-hand side of \eqref{eq:FinalePreTHooft} are determined by the same equation \eqref{sadd}. This gives the functional relation $a \left( \Lambda, \lambda \right)$, but for us it will be convenient to rather invert this to $\lambda \left( \Lambda, a \right)$ and keep $a$ explicitly. Keeping this in mind, the 't Hooft limit of the left-hand side of \eqref{eq:FinalePreTHooft} leads eventually to 
\begin{multline}
\label{eq:SPExpGaugeTheory}
    \exp\left\{ \frac{1}{\epsilon} \mathcal{W}_0^{\mathrm{II}}\left(y\right) + \left[ - \frac{\left(\partial_a{\mathcal W}_0^{\mathrm{II}}\left(y\right)\right)^2}{2 \partial_a^2 \mathcal{F}_0} + \mathcal{W}_1^{\mathrm{II}}\left(y\right) \right] + \mathcal{O}\left(\epsilon\right)\right\} +
    \\ 
    \exp\left\{ \frac{1}{\epsilon} \left[\mathrm{i}\pi \lambda + \mathcal{W}_0^{\mathrm{II}} \left(-y\right) \right] + \left[ \mathrm{i}\partial_y \mathcal{W}_0^{\mathrm{II}}\left(-y\right) - \frac{\left(\partial_a{\mathcal W}_0^{\mathrm{II}}\left(-y\right)\right)^2}{2 \partial_a^2 \mathcal{F}_0} + \mathcal{W}_1^{\mathrm{II}}\left(-y\right) \right] + \mathcal{O}\left(\epsilon\right)\right\},
\end{multline}
where we suppressed the functional dependence on $\Lambda$ and $a$ in the notation.

\subsubsection{The 't Hooft limit on the matrix model side}\label{sec:642}

Consider the Fourier transform on the right-hand side in \eqref{eq:FinalePreTHooft},
\begin{equation}
\label{eq:InverseFourierMM}
    \frac{\sqrt{2 \Lambda}}{\sqrt{2 \pi \epsilon}} \int_\mathbb{R} \mathrm{d}x \exp\left(- \frac{\mathrm{i}}{\epsilon} x y \right) \frac{\exp\left(- 4 \left(\frac{\Lambda}{\epsilon}\right) \cosh\left(x\right) + \frac{x}{2}\right)\Psi_N \left( \mathrm{e}^x, \left(\frac{\Lambda}{\epsilon}\right)^4  \right)}{Z\left( t , N \right)},
\end{equation}
where the 't Hooft expansion of the integrand is
\begin{equation}
    \frac{\exp\left(- 4 \left(\frac{\Lambda}{\epsilon}\right) \cosh\left(x\right) + \frac{x}{2}\right)\Psi_N \left( \mathrm{e}^x, \left(\frac{\Lambda}{\epsilon}\right)^4 \right)}{Z\left( t , N \right)} = \exp\left[ \frac{1}{\epsilon}\mathcal{T}_0\left(\mathrm{e}^x\right) + \frac{x}{2} + \mathcal{T}_1\left(\mathrm{e}^x\right) + \mathcal{O}\left( \epsilon \right)\right] \, ,
\end{equation}
by using \eqref{t2} and again suppressing the functional dependence on the other variables.
In the limit $\epsilon \to 0$, the Fourier transform in \eqref{eq:InverseFourierMM} is dominated by its saddles and becomes
\begin{multline}
\label{eq:MatrixModelTHooftExp}
    \sum_{s} \exp \left\{ \frac{1}{\epsilon} \widehat{\mathcal{T}_{s, 0}} \left(y\right) + \widehat{\mathcal{T}_{s, 1}} \left(y\right)  + \mathcal{O}\left(\epsilon\right) \right\} =
    \\
    \sum_{s} \exp \left\{ \frac{1}{\epsilon} \left[ - \mathrm{i} x_s y + \mathcal{T}_0 \left(\mathrm{e}^{x_s} \right)\right] + \left[ \frac{\log \left( 2 \Lambda \right)}{2} - \frac{\log \left( - \partial_x^2 \mathcal{T}_0 \left(\mathrm{e}^{x_s}\right)\right)}{2} + \frac{x_s}{2} +  \mathcal{T}_1 \left(\mathrm{e}^{x_s}\right) \right] + \mathcal{O}\left(\epsilon\right) \right\}.
\end{multline}
The sum over $s$ is a sum over the saddles and $x_s = x_s\left(y\right)$ is determined by the saddle point equation,
\begin{equation}
\label{eq:SPEqFourietTransTypeI&II}
    y + \mathrm{i} \partial_x \mathcal{T}_0 \left( \mathrm{e}^x \right) = y - \mathrm{i} 2\Lambda \frac{\sqrt{ \mathrm{e}^{2x} -\tg^{2}}\sqrt{\mathrm{e}^{2x}-\tg^{-2}}}{\mathrm{e}^{x}} = 0 \, ,
\end{equation}
where we used \eqref{fin} and $z = \exp \left( x \right)$.
Taking the square of this equation gives the Seiberg-Witten curve \eqref{SWc} if we take as before \eqref{gu},
\begin{equation}
\label{eq:guRepeated}
    \tg^2+\tg^{-2}={u\over 4 \Lambda^2} \, , 
    \qquad \qquad \qquad
    \tg^{\pm 2} = \left(\frac{u}{8 \Lambda^2}\right) \mp \sqrt{\left(\frac{u}{8 \Lambda^2}\right)^2 - 1} \, .
\end{equation}
This leads to the following two solutions,
\begin{equation}
\label{eq:SPESolutions}
    \mathrm{e}^{2x_\pm\left( y \right)} = z_\pm^2 \left(y\right) =  {-\left[\frac{\left( y^2 - u \right) \pm \sqrt{\left( y^2 - u \right)^2 - 64 \Lambda^4}}{8 \Lambda^2}\right]}.
\end{equation}

Let us take a moment to consider the behaviour of $z_\pm\left(y\right)$ as a function of $y$. One can check that $z_\pm\left(y\right)$ is real and outside the branch cut region of the matrix model if and only if $ \mathrm{i} y \in \mathbb{R} \setminus \left\{ 0 \right\}$. It is important to note that with this choice of  $ \mathrm{i} y \in \mathbb{R} \setminus \left\{ 0 \right\}$ one has $z_-^2\left( y \right) > 1 / \tg^2$ and $0 \leqslant z_+^2\left( y \right) < \tg^2$. Moreover, there are no possible choices of $y \in \mathbb{C}$ such that $0 \leqslant z_-^2\left( y \right) < \tg^2$ or $ z_+^2\left( y \right) > 1 / \tg^2$. One finds on the other hand that $z_\pm\left(y\right)$ is real and inside the branch cut region if and only if $0 \leqslant y^2 \leqslant u - 8 \Lambda^2$, and also that $z_\pm\left(y\right)$ is purely imaginary if and only if $y^2 \geqslant u + 8 \Lambda^2$. For all other choices of $y \in \mathbb{C}$, one will find generic complex $z_\pm\left(y\right)$.

\subsubsection{Comparing the gauge theory and the matrix model}
\label{subsubsection:LOTHooft}

To analyze the leading-order of the 't Hooft expansion in \eqref{eq:MatrixModelTHooftExp} with the saddles \eqref{eq:SPESolutions}, it is convenient to separately look at the case $y = 0$ and the derivative with respect to $y$. The reason is that the latter simplifies considerably as a consequence of the saddle point equation \eqref{eq:SPEqFourietTransTypeI&II}. Setting $y = 0$ serves then as a check of the constant term.

Let us first look at the $y$ derivative of the leading-order part. At the matrix model side \eqref{eq:MatrixModelTHooftExp}, one gets by making use of the saddle point equation \eqref{eq:SPEqFourietTransTypeI&II} and its solutions
\be
\label{eq:FourierTransTypeILO}
    \frac{\mathrm{d}}{\mathrm{d}y} \widehat{\mathcal{T}_{\pm, 0}}\left(y\right)
    = - \mathrm{i} x_\pm \left( y \right) \, .
\ee
Comparing this to the leading-order of the gauge theory \eqref{eq:TypeIIDefectEpsilonExp},
we can check that\footnote{Upon using the Matone relation \eqref{mat} and careful treatment of the branches.}
\begin{equation}
\label{eq:TypeIvsIILODelta}
    \begin{split}
        \frac{\mathrm{d}}{\mathrm{d}y} \left[\mathcal{W}_0^{\mathrm{II}}\left(\pm y\right) - \widehat{\mathcal{T}_{\pm S, 0}}\left(y\right) \right]
        =0 \, , 
    \end{split}
\end{equation}
where $S = \mathrm{sgn}\left[\mathrm{arg}\left( \mathrm{i} \left( y^2 - a^2 \right)\right)\right]$ with the convention that $\mathrm{sgn}\left( 0 \right) = -1 \, $.

Let us then look at the constant term for $y = 0$. At the gauge theory side, we have for the leading-order \eqref{eq:TypeIIDefectEpsilonExp}
\begin{equation}
\label{eq:LOTHoofty=0GaugeTheory}
    \mathcal{W}_0^{\mathrm{II}}\left( 0 \right)  
    = - \frac{\pi}{2} \left|a\right|.
\end{equation}
From \eqref{eq:SPESolutions}, one can see that  $z_\pm\left( 0 \right) = \tg^{\pm} > 0$, with $\tg^{\pm}$ as in \eqref{eq:guRepeated}. Using \eqref{eq:def:MatrixModelT0} gives for the leading-order of the matrix model \eqref{eq:MatrixModelTHooftExp}
\begin{equation}
    \widehat{\mathcal{T}_{\pm, 0}} \left( 0 \right)
    =
    \mathcal{T}_0 \left(z_\pm\left(0\right)\right)
    =
    - 2 \Lambda \left( \tg + \tg^{-1} \right) + 2 \int_{+ \infty}^{\tg^{\pm 1}} \mathrm{d} z \, \lambda \omega_+^0\left(z\right) \, ,
\end{equation}
where the even planar resolvent $\omega_+^0\left(z\right)$ is given in \eqref{planar4d}. Note that the difference between the leading terms of the two saddles is
\begin{equation}
\label{eq:LOTHoofty=0MMDiff}
    \left[ \widehat{\mathcal{T}_{+, 0}} \left(0\right) - \widehat{\mathcal{T}_{-, 0}} \left(0\right) \right]
    =
    - 2 \int_{\tg}^{\tg^{-1}} \mathrm{d} z \, \lambda \omega_+^0\left(z\right) =  \mathrm{i} \pi \lambda \, .
\end{equation}
The last equality can be obtained in an $\Lambda \to 0$ expansion or exactly using \cite[eq.~4.18]{Eynard:1995nv}, which shows that this relation does not depend on the particular form of the potential. We have furthermore that
\begin{multline}
\label{eq:LOTHoofty=0MMT0Minus}
    \widehat{\mathcal{T}_{-, 0}} \left( 0 \right)
    =
    - 2 \Lambda \left( \tg + \tg^{-1} \right) + 2 \int_{+ \infty}^{\tg^{- 1}} \mathrm{d} z \ \lambda \omega_+^0\left(z\right)
    \\
    = - 2 \Lambda \left[\frac{2 \mathrm{E}\left( \tg^4 \right) + \left( \tg^{-2} - \tg^2 \right) \left[ - 2 \tg^2 \mathrm{K}\left( \tg^4 \right) + \mathrm{K} \left( \tg^{-4} \right) + \mathrm{i} \mathrm{K}\left( 1 -\tg^{-4} \right) \right]}{\tg}\right] = - \frac{\pi}{2} \left| a \right| \, ,
\end{multline}
with $\mathrm{K}$ and $\mathrm{E}$ the complete elliptic integrals of the first \eqref{eq:EllipticF} and second \eqref{eq:EllipticE} kind, respectively. The last equality was found in an $\Lambda \to 0$ or equivalently $\tg \to 0$ expansion using \eqref{eq:guRepeated} and the Matone relation \eqref{mat}.
Hence, from \eqref{eq:LOTHoofty=0GaugeTheory}, \eqref{eq:LOTHoofty=0MMDiff}, and \eqref{eq:LOTHoofty=0MMT0Minus}
\begin{equation}
    \begin{split}
        \mathcal{W}_0^{\mathrm{II}}\left(0\right) - \widehat{\mathcal{T}_{-, 0}}\left(0\right)
        &
        = 0 \, ,
        \\
        \mathrm{i} \pi \lambda + \mathcal{W}_0^{\mathrm{II}}\left( 0 \right) - \widehat{\mathcal{T}_{+, 0}}\left( 0 \right)
        &
        = 0 \, .
    \end{split}
\end{equation}
So the constant parts of the leading-order terms agree, and together with \eqref{eq:TypeIvsIILODelta} this proves the equality in \eqref{eq:FinalePreTHooft} and hence \eqref{finalecompact} to leading-order in the 't Hooft limit \eqref{eq:THooftScalingNEpsilon}.

The subleading-order can be checked in analogy with \autoref{TypeIDefect}. Matching at higher order in $\epsilon$ can then be inferred from topological recursion, as we discussed near the end of \autoref{TypeIDefect}.

\subsection{Numerical eigenfunctions}\label{sec:numeig}

The numerical analysis of the spectrum and the eigenfunctions for the integral kernel $\rho$ \eqref{kernrho} is done exactly as in \cite[sec.~2.2]{Hatsuda:2012hm}. To make the presentation self-contained, let us review the strategy of \cite[sec.~2.2]{Hatsuda:2012hm}. We are interested in studying numerically the eigenvalue equation
\be \label{eigennu}\int_{\IR}\rd y \, \rho(x, y)\varphi_n(y,t)=E_n \varphi_n(x,t) \, , \ee
where the kernel $\rho(x,y)$ is defined in \eqref{kernrho}.
It is  convenient to decompose $\rho(x,y)$  as
\be \rho(x,y)=\sum_{k \geqslant 0}\rho_k(x)\rho_k(y) \, ,
\qquad \qquad \qquad \rho_k(x)=\frac{\tanh^k\left(\frac{x}{2}\right) \exp \left(-4 t^{1/4} \cosh (x)\right)}{\cosh \left(\frac{x}{2}\right)} \, ,
\ee
and to define
\be v_k^{(n)}(t)=\int_{\IR}\rd y \, \rho_k(y) \varphi_n(y,t) \, . \ee
Then, \eqref{eigennu} reads
\be \label{vrho}\sum_{k\geqslant 0} \rho_k(x) v_k^{(n)}(t) =E_n \varphi_n(x,t) \, , \ee
which we can also write as
\be\sum_{k\geqslant 0}H_{\ell, k}v_k^{(n)}(t)=E_n v_{\ell}^{(n)}(t)~, \ee
with $H$ the infinite-dimensional Hankel matrix defined by
\be \label{hankel} H_{k, \ell}=\int_{\mathbb{R}} \rd x \frac{\tanh ^{k + \ell}\left(\frac{x}{2}\right) \exp \left(-8 {t^{1/4}} \cosh (x)\right)}{\cosh ^2\left(\frac{x}{2}\right)} \, ,
\qquad \qquad
k , \ell \geqslant 0 \, .
\ee
This means that the eigenvalues of $H$ coincide with those of 
$\rho \left( x, y \right)$ and the eigenvectors of $H$ give the eigenfunctions of $\rho \left( x, y \right)$ via \eqref{vrho}. The advantage of working with $H$ is that we can numerically compute its eigenvalues and eigenfunctions by truncating the matrix to a finite size while maintaining control over the numerical error due to the truncation. 
Let $v^{(n,M)}(t)$ be the $n^{\rm th}$ eigenvector of the Hankel matrix $H$ \eqref{hankel} truncated at size $M$. Defining $\varphi_n^{(M)}(x,t)$ by
\be
\varphi_n^{(M)}(x,t) = \sum_{k = 0}^M \rho_k(x) v^{(n,M)}_k \, , 
\qquad \qquad
v^{(n,M)}=\left(\begin{array}{c}
     v^{(n,M)}_0  \\
     \vdots \\
     v^{(n,M)}_M
\end{array}\right) \, ,
\ee 
we recover the true eigenfunctions of the kernel $\rho$ in the $M \to + \infty$ limit,
\be \label{eigenume}\lim_{M \to + \infty}\varphi_n^{(M)}(x,t)\propto \varphi_n(x,t) \, , \ee
where the proportionality factor is a numerical constant and $\varphi_n$ is the $n^{\rm th}$ eigenfunction of \eqref{eigennu} in the normalization of \eqref{eigenfff}. 
We computed the left-hand side of \eqref{eigenume} numerically and checked that this numerical expression agrees with the eigenfunctions computed by using the defect expression on the right-hand side of \eqref{eigenfff} with high precision. For instance, for $t=1/(100 \pi^8) $, by including $0$ instantons in  \eqref{eigenfff} we get a pointwise agreement  of the order $10^{-6}$. Likewise, by including 1, 2 and 3 instantons we get a pointwise agreement of the order $10^{-11}$, $10^{-16}$, and   $10^{-22}$, respectively\footnote{One can reach this precision with Hankel matrices of size $M = 2^{9} = 512$.}.

\section{Outlook}\label{outlook}
 
In this paper, we have shown that the eigenfunctions of the operator \eqref{ointro} are computed by surface defects in $\mathcal{N}=2$, $\su2$  SYM  in the self-dual phase of the four-dimensional $\Omega$-background ($\epsilon_1+\epsilon_2=0$). This result, together with \cite{bgt, bgt2,Gavrylenko:2023ewx}, extends the correspondence between 4d $\mathcal{N}=2$ theories and spectral theory to a new class of operators.
In addition, we have expressed the eigenfunctions of these operators in closed form via a matrix model average \eqref{psiNintro}.  
This provides a representation for the surface defect partition function which resums both the instanton and the $\epsilon$ expansions.
In this way, we have a manifest interpolation from the weak to the strong coupling region.  In particular,  the strong coupling expansion in $1/\Lambda$ (exact in $\epsilon$ and $a_D$) corresponds to the expansion of the matrix model around its Gaussian point, and hence, it is obtained straightforwardly. 

Some further comments and generalizations:
\begin{itemize}
\item[-] In this work, we focused on the specific example of 4d, $\mathcal{N}=2$, $\su2$ SYM and the operator \eqref{ointro}. It would be interesting to extend our results in a systematic way to all 4d $\mathcal{N}=2$ theories. For example, in the case of $\mathcal{N}=2$, $\suN$  SYM we have $N-1$ non-commuting  Fermi gas operators as discussed in \cite{bgt2,Gavrylenko:2023ewx}. We expect their eigenfunctions to be computed by surface defects in $\suN$ SYM in the self-dual phase of the $\Omega$ background.

\item[-] Our results should follow from the {\it open} version of the TS/ST correspondence \cite{Marino:2016rsq,Marino:2017gyg}. We will report on this somewhere else \cite{matijn-alba}. 

\item[-] The Fredholm determinant of \eqref{ointro} computes the tau function of the Painlev\'e $\rm III_3$ equation at specific initial conditions. It would be interesting to understand the role of the eigenfunctions of \eqref{ointro} in the context of Painlev\'e equations. In particular, the relation to the solution of the linear system associated with Painlev\'e equations \cite{Iorgov:2014vla,Jeong:2021rll} as well as with \cite{Bonelli:2022iob}.

\item[-] The Fredholm determinant and the spectral traces of \eqref{ointro} can also be expressed via a pair of coupled TBA equations closely related to \emph{two-dimensional}  theories  \cite{Cecotti:1992qh, zamo}. It would be interesting to understand this better since this may reveal an interesting 4d-2d interplay characterizing directly the self-dual phase of the $\Omega$-background.

\item[-] The operator \eqref{ointro} is a particular example of a Painlev\'e kernel whose Fredholm determinant computes the tau function. A more general class of Fredholm determinants was constructed in \cite{Gavrylenko:2016zlf,Gavrylenko:2017lqz,Desiraju2020FredholmDR,DelMonte:2020wty}. It would be interesting to see if also in this case the corresponding (formal) eigenfunctions are related to surface defects. 

\item[-] It is well known that the canonical quantization of the SW curve for $\su2$ SYM leads to the Mathieu operator \eqref{mathieu}. We expect a different quantization scheme to produce the operator  \eqref{ointro}. It is important to understand what this other quantization scheme is. Since the spectral analysis of \eqref{ointro} is encoded in the self-dual phase of the $\Omega$-background, a natural quantization scheme to investigate would be the one arising in the context of the topological recursion \cite{Iwaki:2015xnr,Iwaki:2019zeq,Marchal:2019nsq,Marchal:2019bia}.

\end{itemize}

\section*{Acknowledgments}

We would like to thank Anton Nedelin for collaborating during the early stages of this work. We are grateful to Severin Charbonnier, Shi Cheng, Qianyu Hao, Saebyeok Jeong, Rinat Kashaev, Marcos Mari\~no, Andy Neitzke, Nicolas Orantin, François Pagano, Sara Pasquetti, Lukas Schimmer, Yasuhiko Yamada, and Szabolcs Zakany for useful discussions and correspondence. Furthermore, we thank the two anonymous referees for their comments and valuable suggestions, which led to clarifications in the second, revised version of the paper.
This work is partially supported by the Swiss National Science Foundation Grant No. 185723 and the NCCR SwissMAP.

\hfill

\noindent
This version of the article has been accepted for publication, after peer review, but is not the Version of Record. The Version of Record of this article is published in Annales Henri Poincar\'e, and is available online at: \url{https://doi.org/10.1007/s00023-024-01469-4}.

\section*{Declarations}

\subsection*{Conflicts of interest}

The authors declare that they have no conflicts of interest to disclose.

\appendix
 
\section{The refined open topological vertex and the type II defect}\label{app:vertex}

In this appendix, we explain how the type II defect partition function of 4d, $\mathcal{N} = 2$, $\mathrm{SU}\left(2\right)$ super Yang-Mills can be obtained from open topological strings on the toric Calabi-Yau manifold known as local $\mathbb{F}_0 = \mathbb{P}^1 \times \mathbb{P}^1$.

\subsection{Open topological strings on local \texorpdfstring{$\mathbb{F}_0$}{F0}}

To find the open string partition function on local $\mathbb{F}_0$, we use \cite{Iqbal:2007ii} and \cite{Cheng:2021nex}.  We will closely follow the notation and presentation of \cite{Cheng:2021nex}.

\subsubsection{Notation and conventions}

The variables $Q_{B,F}$ are related to the Kähler parameters $T_{B,F}$ of the base and fiber $\mathbb{P}^1$'s, respectively,
\begin{equation}
    Q_{B,F} =\exp \left( - T_{B,F} \right) \, ,
    \qquad \qquad
    T_{B,F} = - \log \left( Q_{B,F} \right) \, , 
\end{equation}
and $Q_1$ is similarly the open Kähler parameter of the brane \cite[pp.~9-14,~50]{Cheng:2021nex}.
The variables $t, q$ are similarly connected to the $\Omega$-background parameters $\epsilon_{1,2}$ by 
\begin{equation}
    t =\exp \left( \mathrm{i} \epsilon_1 \right) \, , \qquad \qquad \qquad q =\exp \left( - \mathrm{i} \epsilon_2 \right) \, .
\end{equation}
It is important to note that these variables are not related to the $t$ and $q$ appearing in the main text.

\subsubsection{The refined open topological vertex}

Important building blocks of the partition function are the so-called Nekrasov factors
\begin{equation}
    \begin{split}
        N_{\mu, \nu} \left( Q; t, q \right) & =\prod_{k, \ell = 1}^{+ \infty} \frac{1 - Q \ q^{\nu_k - \ell} t^{\mu_\ell^t-k+1}}{1 - Q \ q^{-\ell} t^{-k+1}} \\
        & = \prod_{\left(k, \ell\right) \in \nu} \left( 1 - Q \ q^{\nu_k - \ell} t^{\mu_\ell^t-k+1} \right) \prod_{\left(k, \ell\right) \in \mu} \left( 1 - Q \ q^{-\mu_k + \ell - 1} t^{-\nu_\ell^t+k} \right) \, ,
    \end{split}
\end{equation}
where $\mu$ and $\nu$ are integer partitions:
\begin{equation}
    \mu = \left\{ \mu_1, \mu_2, \mu_3, \cdots \ | \ \forall \ k, \ell \in \mathbb{N} \setminus \left\{ 0 \right\}: \left( \mu_k \in \mathbb{N} \right) \land \left(  k \leqslant \ell \Rightarrow \mu_k \geqslant \mu_\ell \right) \right\} \, .
\end{equation}
Hence, $\left(k, \ell \right) \in \mu$ is any pair $k, \ell \in \mathbb{N}\setminus \left\{ 0 \right\}$ such that $1 \leqslant \ell \leqslant \mu_k$. Integer partitions can be represented by a Young diagram, and the transposed partition $\mu^t$ is obtained by transposing the Young diagram,
\begin{equation}
    \mu^t =\left\{ \mu_1^t, \mu_2^t, \mu_3^t, \cdots \ | \ \forall \ \ell \in \mathbb{N} \setminus \left\{ 0 \right\}: \mu_\ell^t =\left| \left\{ k \in \mathbb{N} \setminus \left\{ 0 \right\} \ | \ \mu_k \geqslant \ell \right\} \right|\right\} \, .
\end{equation}
The key ingredient in the topological vertex computation is\footnote{The first Nekrasov factor on the second line of \eqref{eq:NekrasovFactors} differs from what is given in \cite[p.~50, eq.~5.4]{Cheng:2021nex} where they have $Q_F \leftrightarrow Q_F q$. However, comparison with \cite[p.~35, eq.~90]{Iqbal:2007ii} and \cite[p.~27, eqs.~4.9,~4.11]{Marino:2017gyg} leads us to \eqref{eq:NekrasovFactors}.},
\begin{multline}
\label{eq:NekrasovFactors}
        C_{\mu, \nu} \left( Q_1, Q_F, t, q \right) =\frac{N_{\emptyset, \mu^t} \left( Q_1 \frac{t^2}{q}; t^{-1}, q^{-1} \right) N_{\emptyset, \nu} \left( Q_1 Q_F \frac{t^2}{q}; t^{-1}, q^{-1} \right)}{N_{\emptyset, \mu^t} \left( Q_1 \frac{t}{q}; t^{-1}, q^{-1} \right) N_{\emptyset, \nu} \left( Q_1 Q_F \frac{t}{q}; t^{-1}, q^{-1} \right)}
        \\
        \frac{1}{N_{\mu^t,\nu}\left( Q_F; t^{-1}, q^{-1} \right) N_{\mu^t,\nu}\left( Q_F \frac{t}{q}; t^{-1}, q^{-1} \right)} \, ,
\end{multline}
where $\emptyset$ is the empty partition $\{\}$.
This gives for the ``\textit{open-closed t-brane partition function}" of local $\mathbb{F}_0$ \cite[p.~50, eq.~5.4]{Cheng:2021nex}
\begin{multline}
    Z^{\text{open-closed, t-brane}}_{\mathbb{F}_0} \left(  Q_1 , Q_B, Q_F, t, q \right) =
    \\
    \sum_{\mu, \nu} Q_B^{\left| \mu \right| + \left| \nu \right|} t^{||\nu^t||^2} q^{||\mu^t||^2}
    ||Z_\mu \left( t, q \right)||^2 ||Z_\nu \left( q, t \right)||^2 C_{\mu, \nu} \left(  Q_1 , Q_F, t, q \right) \, ,
\end{multline}
where one defines 
\begin{equation}
    |\mu| =\sum_{k=1}^{+\infty} \mu_k \, , \qquad \qquad ||\mu||^2 =\sum_{k=1}^{+\infty} \mu_k^2 \, ,
\end{equation}
\begin{equation}
\label{eq:defZtilde3}
    \left| \left| Z_\mu \left( t, q \right) \right| \right|^2 =Z_{\mu^t} \left( t, q \right) Z_\mu \left( q, t \right) \, ,
    \qquad \qquad
    Z_\mu \left( t, q \right) =\prod_{\left( k , \ell \right) \in \mu} \left( 1 - q^{\mu_k - \ell} t^{\mu_\ell^t - k + 1} \right)^{-1} \, .
\end{equation}
The t-brane instanton partition function for local $\mathbb{F}_0$ is then \cite[p.~50, eq.~5.7]{Cheng:2021nex}
\begin{equation}
\label{eq:T-BraneInstanton}
        Z^{\text{t-brane, instanton}}_{\mathbb{F}_0} \left(  Q_1 , Q_B, Q_F , t, q \right) =\frac{Z^{\text{open-closed, t-brane}}_{\mathbb{F}_0} \left(  Q_1 , Q_B, Q_F, t, q \right)}{Z^{\text{closed}}_{\mathbb{F}_0} \left( Q_B, Q_F, t, q \right)} \, ,
\end{equation}
where $Z^{\text{closed}}_{\mathbb{F}_0}$ is given by
\begin{equation}
    Z^{\text{closed}}_{\mathbb{F}_0} \left( Q_B, Q_F, t, q \right) = Z^{\text{open-closed, t-brane}}_{\mathbb{F}_0} \left( 0 , Q_B, Q_F, t, q \right) \, .
\end{equation}
In addition, one has also the 1-loop part \cite[pp.~50,~53, eqs.~5.6,~A.15,~A.22,~A.25]{Cheng:2021nex}
\begin{equation}
\label{eq:T-Brane1Loop}
    Z^\text{t-brane, 1-loop}_{\mathbb{F}_0}\left(  Q_1 , Q_F, t, q \right) = \left( \frac{Q_1 t}{q} ; \frac{1}{q} \right)_\infty \left( \frac{Q_1 Q_F t}{q} ; \frac{1}{q} \right)_\infty = \frac{1}{\left( Q_1 t ; q \right)_\infty \left( Q_1 Q_F t ; q \right)_\infty } \, ,
\end{equation}
where $\left( a ; q \right)_\infty$ is the $q$-Pochhammer symbol. The complete partition function is then given by
\begin{multline}
\label{eq:T-Brane}
    Z^{\text{t-brane}}_{\mathbb{F}_0} \left(  Q_1 , Q_B, Q_F, t, q \right) =
    \\
    Z^{\text{t-brane, 1-loop}}_{\mathbb{F}_0} \left(  Q_1 , Q_F, t, q \right) Z^{\text{t-brane, instanton}}_{\mathbb{F}_0} \left(  Q_1 , Q_B, Q_F, t, q \right) \, .
\end{multline}
Two particular phases of the refined topological vertex are of interest to us: the self-dual or Gopakumar-Vafa (GV) phase $\epsilon_1 + \epsilon_2 = 0$ or $t = q$, and the Nekrasov-Shatashvili (NS) phase $\epsilon_1 = 0$ or $t = 1$.

\subsubsection{Non-perturbative topological strings}

We obtain the unrefined topological string amplitude if we set $g_s = - \epsilon_1 = \epsilon_2$.
In addition, it was found in \cite{hmmo, ghm, cgm2}  that when one includes the NS phase of the refined topological vertex as well one obtains a non-perturbative completion of the topological string partition function on toric CYs. This was generalized to some extent from the closed to the open vertex in \cite{Marino:2016rsq, Marino:2017gyg},
see also \cite{Kashani-Poor:2016edc, Sciarappa:2017hds}. 
In particular, for local $\IF_0$ one finds \cite{Marino:2016rsq, Marino:2017gyg}
\begin{equation}
\label{eq:ZopenF0}
    Z_{\mathbb{F}_0}^\text{open, np}\left( X, Q_B, Q_F, g_s \right) = Z_{\mathbb{F}_0}^\text{(p)}\left( X, g_s \right) Z_{\mathbb{F}_0}^\text{GV}\left( X, Q_B, Q_F, g_s \right) Z_{\mathbb{F}_0}^\text{NS}\left( X^S, Q_B^S, Q_F^S, g_s \right) \, .
\end{equation}
We explain all functions and symbols involved below. The usual open topological string part $Z_{\mathbb{F}_0}^\text{GV}\left( X, Q_B, Q_F, g_s \right)$ is the self-dual phase of the open refined topological vertex \eqref{eq:T-BraneInstanton}, \eqref{eq:T-Brane1Loop}, \eqref{eq:T-Brane},
\begin{equation}
\label{eq:ZGV}
    \begin{split}
        Z_{\mathbb{F}_0}^\text{GV}\left( X, Q_B, Q_F, g_s \right)  = Z^{\text{t-brane}}_{\mathbb{F}_0} \left( \frac{\sqrt{q}}{\sqrt{Q_F} X}, Q_B, Q_F, \frac{1}{q}, \frac{1}{q} \right) \, ,
        \quad \qquad
        q  = \re^{\mathrm{i} g_s} \, .
    \end{split}
\end{equation}
On the other hand, one has $Z_{\mathbb{F}_0}^\text{NS}\left( X^S, Q_B^S, Q_F^S, g_s \right)$ which is the NS phase of the refined topological vertex, \eqref{eq:T-BraneInstanton}, \eqref{eq:T-Brane1Loop}, \eqref{eq:T-Brane},
\begin{equation}
\label{eq:ZNS}
    Z_{\mathbb{F}_0}^\text{NS}\left( X^S, Q_B^S, Q_F^S, g_s \right)
    = \lim_{p^S \to 1} Z^{\text{t-brane}}_{\mathbb{F}_0} \left( - \frac{\sqrt{p^S} }{\sqrt{Q_F^S} X^S}, Q_B^S, Q_F^S, \frac{1}{p^S}, \frac{1}{q^S} \right) \, ,
    \qquad 
    q^S
    = \re^{\mathrm{i} \frac{\left( 2 \pi \right)^2}{g_s}} \, ,
\end{equation}
which is perturbative in the inverse string coupling $g_s^{-1}$. The exponentiated Kähler parameters $X, Q_B, Q_F$ and their NS counterparts $X^S, Q_B^S, Q_F^S$ are related by powers of $2 \pi / g_s$,
\begin{equation}
    A^S =A^{2 \pi / g_s} \, , \qquad \qquad \qquad A = \left(A^S\right)^{g_s / 2 \pi} \, ,
\end{equation}
where $A$ stands for either one of $X, Q_B, Q_F$.
The ``polynomial" part $Z^\text{(p)}\left( X, g_s \right)$ is not obtained from the topological vertex but is given by \cite[eq.~4.10]{Marino:2017gyg}\footnote{In \cite{Marino:2017gyg}, they focus on the special case where there is only a single Kähler parameter, so $Q_B = Q_F$. One can check, however, that \cite[eq.~4.10]{Marino:2017gyg} is also correct for $Q_B \neq Q_F$. To get from \cite[eq.~4.10]{Marino:2017gyg} to \eqref{jp}, one needs to take $x = \log \left(X^\mathrm{S}\right)$ and $\hbar = \left( 2 \pi \right)^2 / g_s$.}
\begin{equation}
\label{jp}
    \begin{split}
        Z_{\mathbb{F}_0}^\text{(p)} \left( X, g_s \right)
        & = \exp \left\{- \frac{\mathrm{i}}{2} \frac{1}{g_s} \left[ \log^2 \left( X \right) - \mathrm{i} 2 \pi \log \left( X \right) \right] + \frac{\log \left( X \right)}{2} \right\}
        \\
        & = \exp\left\{- \frac{\mathrm{i}}{2} \frac{g_s}{\left( 2 \pi \right)^2} \left[ \log^2 \left( X^S \right) + \mathrm{i} 2 \pi \log \left( X^S \right) \right] -  \frac{\log \left( X^S \right)}{2} \right\} \, ,
    \end{split}
\end{equation}
which explains all of the functions and symbols appearing in \eqref{eq:ZopenF0}.

One can see that the GV and NS instanton parts are given as series expansions in $Q_B$ and $Q_B^S$, respectively,
\begin{equation}
\label{eq:ZInstantonF0}
    \begin{split}
        & Z_{\mathbb{F}_0}^\text{GV, inst} \left( X, Q_B, Q_F, g_s \right) = 
        \\
        & \ \
        1 - \left[\frac{ Q_1 \left( 2 Q_F Q_1 - Q_F - 1 \right)}{\left( q - 1 \right) \left( Q_F - 1 \right)^2 \left( Q_1 - 1 \right) \left( Q_F Q_1 - 1 \right)}\right] Q_B + \mathcal{O}\left(Q_B^2\right)
        \\
        & Z_{\mathbb{F}_0}^\text{NS, inst}\left( X^S, Q_B^S, Q_F^S, g_s \right) =
        \\
        & \ \
        1 - \left[\frac{\left(q^S\right)^2 Q_1^S \left( q^S \left(q^S + 1 \right)Q_F^S Q_1^S - Q_F^S - 1\right)}{(q^S-1) (Q_F^S - q^S) (q^S Q_F^S-1) (q^S Q_1^S-1) (q^S Q_F^S Q_1^S-1)}\right] Q_B^S + \mathcal{O} \left(\left(Q_B^S\right)^2\right)
    \end{split}
\end{equation}
where $Q_1 = \sqrt{q} / \sqrt{Q_F} X$ and  $Q_1^S = - 1 / \sqrt{Q_F^S} X^S$. 

By looking at the full expression \eqref{eq:ZopenF0}, we see that two distinct four-dimensional limits can be implemented: one in which $Z^{\rm GV}_{\IF_0}$ survives, and another in which $Z^{\rm NS}_{\IF_0}$ survives. Using the TS/ST correspondence \cite{ghm} one can show that at the level of the operator theory the first limit makes contact with \eqref{dual4d} and \eqref{limitdual4do}. The second limit instead gives \eqref{limit1} and \eqref{limit1op}. A more detailed explanation of this phenomenon can be found in \cite{bgt, bgt2, Gavrylenko:2023ewx}. In this paper, our primary focus is directed towards the first limit.

\subsubsection{The 1-loop part in terms of Faddeev's quantum dilogarithm}

To take the 4d limit of the partition function \eqref{eq:ZopenF0}, we first need to rewrite the 1-loop part and combine $Z_{\mathbb{F}_0}^\text{GV, 1-loop}$ and $Z_{\mathbb{F}_0}^\text{NS, 1-loop}$ into a single function. From \eqref{eq:T-Brane1Loop}, \eqref{eq:ZGV} and \eqref{eq:ZNS} we have
\begin{equation}\label{1loops}
    \begin{split}
        Z^\text{GV, 1-loop}_{\mathbb{F}_0}\left( X, Q_F, q  \right) & = \left(\frac{\sqrt{q}}{\sqrt{Q_F} X} ; q \right)_\infty \left( \frac{\sqrt{q} \sqrt{Q_F}}{X} ; q \right)_\infty
        \\
        & = \left[ \left( \frac{1}{\sqrt{q} \sqrt{Q_F} X} ; \frac{1}{q} \right)_\infty \left( \frac{\sqrt{Q_F}}{\sqrt{q} X} ; \frac{1}{q} \right)_\infty \right]^{-1},
        \\
        Z^\text{NS, 1-loop}_{\mathbb{F}_0}\left( X^S, Q_F^S, q^S  \right) & = \left( - \frac{q^S}{\sqrt{Q_F^S} X^S} ; q^S \right)_\infty \left( - \frac{q^S \sqrt{Q_F^S}}{X^S} ; q^S \right)_\infty
        \\
        & = \left[ \left( - \frac{1}{\sqrt{Q_F^S} X^S} ; \frac{1}{q^S} \right)_\infty \left( - \frac{\sqrt{Q_F^S}}{X^S} ; \frac{1}{q^S} \right)_\infty \right]^{-1},
    \end{split}
\end{equation}
where $\left( a ; q \right)_\infty$ is the $q$-Pochhammer symbol. The product of $Z^\text{GV, 1-loop}_{\mathbb{F}_0}$ and $Z^\text{NS, 1-loop}_{\mathbb{F}_0}$ can then be written as a product of quantum dilogarithms.

Faddeev's quantum dilogarithm can be defined by \cite{faddeev} \cite[eq.~2]{Kashaev:2000ku} \cite[eqs.~157,~158]{kama}
\begin{equation}
    \Phi_b \left(z\right) = \frac{\left(-\mathrm{e}^{\ri \pi b^2 + 2\pi b z}; \mathrm{e}^{\ri 2 \pi b^2}\right)_\infty}{\left(-\mathrm{e}^{- \ri \pi b^{-2} + 2 \pi b^{-1} z} ; \mathrm{e}^{ - \ri 2 \pi b^{-2} }\right)_\infty} \, ,
    \qquad \qquad \qquad
    \mathrm{Im}\left(b^2\right) > 0 \, ,
\end{equation}
and it can be extended to all $b^2 \in \mathbb{C}\setminus\mathbb{R}_{\leqslant 0}$. The quantum dilogarithm\footnote{Faddeev's quantum dilogarithm appears in many fields of science under various names \cite[pp.~5-6]{Faddeev:2014hia}. It is also known as ``non-compact quantum dilogarithm" or simply ``quantum dilogarithm" \cite[p.~2]{Kashaev:2000ku}, and it is closely related to the ``modular quantum dilogarithm" from spectral theory, the ``double sine function" from number theory, and the ``hyperbolic gamma function" from integrable systems \cite[pp.~5-6]{Faddeev:2014hia}.} is a meromorphic function of $z$ with zero's along the negative imaginary axis and poles along the positive imaginary axis when $b > 0$ \cite[eq.~160]{kama}, and with an essential singularity at infinity \cite[p.~2]{Kashaev:2000ku}.
If one takes $b^2 = 2\pi / g_s > 0$, one has
\begin{equation}
    q = \mathrm{e}^{\mathrm{i} g_s} = \mathrm{e}^{\mathrm{i} 2 \pi b^{-2}} \, , 
    \qquad \qquad \qquad
    q^S =  \mathrm{e}^{\mathrm{i} \left( 2 \pi \right)^2 / g_s} = \mathrm{e}^{\mathrm{i} 2 \pi b^2} \, ,
\end{equation}
and defining $z_+$ and $z_-$ by
\begin{equation}
\label{eq:Z1LoopF0Defz}
    \frac{\left(\sqrt{Q_F}\right)^{\mp 1}}{X} = \mathrm{e}^{2 \pi b^{-1} z_\pm - \ri \pi} = \left(\frac{\left(\sqrt{Q_F^S}\right)^{\mp 1}}{X^S} \right)^{1/b^2} \, ,
\end{equation}
one gets for the 1-loop part of the partition function \eqref{eq:ZopenF0}
\begin{equation}
\label{eq:Z1LoopF0}
    Z_{\mathbb{F}_0}^\text{1-loop} = Z^\text{GV, 1-loop}_{\mathbb{F}_0} Z^\text{NS, 1-loop}_{\mathbb{F}_0} = \Phi_b \left(z_+\right)  \Phi_b \left(z_-\right) \, .
\end{equation}
A useful property of Faddeev's quantum dilogarithm $\Phi_b\left(z\right)$ is that it is invariant under inversion of $b$ with $z$ kept fixed,
\begin{equation}
\label{eq:FaddQuantDiLogSDual}
    \Phi_b \left( z \right) = \Phi_{1/b} \left( z \right) \, .
\end{equation}
In a physics language, this corresponds to a sort of S-duality\footnote{See also \cite{Wang:2015wdy,Hatsuda:2015qzx} for a discussion about the $\hbar\leftrightarrow 1/\hbar$ symmetry in the context of the TS/ST correspondence and relativistic integrable systems.}, since it inverts the string coupling $g_s$.

An integral representation of the quantum dilogarithm which will be useful for us is \cite[eq. A.2]{Hatsuda:2016uqa} 
\begin{equation}
\label{eq:FaddeevQuantumDilogarithmDoubleSineReps}
    \begin{split}
        \log \left( \Phi_b\left(z\right) \right) & = - \frac{\mathrm{i}}{2 \pi b^2} \mathrm{Li}_2 \left(-\mathrm{e}^{2\pi b z}\right)  - \mathrm{i} \int_0^{+\infty} \frac{\mathrm{d}t}{1 + \mathrm{e}^{2\pi t}} \log \left(\frac{1 + \mathrm{e}^{2 \pi b z} \mathrm{e}^{- 2 \pi b^2 t}}{1 + \mathrm{e}^{2 \pi b z} \mathrm{e}^{2 \pi b^2 t}}\right)\\
        & = - \frac{\mathrm{i}}{2 \pi} b^2 \mathrm{Li}_2 \left(-\mathrm{e}^{2\pi b^{-1} z }\right)  - \mathrm{i} \int_0^{+\infty} \frac{\mathrm{d}t}{1 + \mathrm{e}^{2\pi t}} \log \left(\frac{1 + \mathrm{e}^{2 \pi b^{-1} z} \mathrm{e}^{- 2 \pi b^{-2} t}}{1 + \mathrm{e}^{2 \pi b^{-1} z} \mathrm{e}^{2 \pi b^{-2} t}}\right) \, ,
    \end{split}
\end{equation}
where the second representation follows from the inversion symmetry of the quantum dilogarithm \eqref{eq:FaddQuantDiLogSDual}. It is this second integral representation which will be useful in taking the 4d limit.

\subsection{A four-dimensional limit}

Consider the following reparametrization of the Kähler parameters \cite[p.~5]{bgt}
\begin{equation}
\label{eq:ParametrizationDual4DLimit}
    X = \exp\left(g_s x \right) \, ,
    \qquad \qquad
    Q_B = g_s^4 t \, ,
    \qquad \qquad
    Q_F = \exp \left( 2 g_s \mathrm{i} \sigma \right) \, ,
\end{equation}
where as before $q = \exp\left( \mathrm{i} g_s \right)$ and $q^S = \exp\left( \mathrm{i} \left( 2 \pi \right)^2 / g_s \right)$.
The four-dimensional limit is then $g_s \to 0$  with $x$, $t$ and $\sigma$ and kept fixed. In terms of more familiar gauge theoretic quantities, one has
\begin{equation}
    x = \frac{y}{2 \epsilon} \, ,
    \qquad \qquad
    t = \left(\frac{\Lambda}{\epsilon}\right)^4 \, ,
    \qquad \qquad
    \sigma = \mathrm{i} \frac{a}{ 2 \epsilon} \, , 
\end{equation}
where $\epsilon $ corresponds to the $\Omega$-background parameter in the self-dual phase, $\Lambda$ is the instanton counting parameter and, $a$ is the Coulomb branch parameter or equivalently the A-period of the underlying Seiberg-Witten geometry. The parameter $y$ corresponds to the position of the defect. Note also that the variable $x$ is called $q$ in the main text.

The instanton part of the type II defect is obtained as the $g_s \to 0$ limit of the GV instanton part \eqref{eq:ZInstantonF0} of the topological string partition function in the parametrization \eqref{eq:ParametrizationDual4DLimit},
\begin{multline}
\label{eq:ZDTypeIIInstanton}
    Z_{\rm inst}^\text{II}\left( x, t, \sigma \right) 
    = \lim_{g_s \to 0} Z_{\mathbb{F}_0}^\text{GV, inst} \left(  \exp \left(g_s x \right), g_s^4 t, \exp\left( 2 g_s \ri \sigma \right), g_s \right)
    = 1 - \left[ \frac{\widetilde{x}}{2 \sigma^2 \left(\widetilde{x}^2 - \sigma^2 \right)}\right] t
    \\
   + \left[ \frac{\widetilde{x} \left( \widetilde{x} + 1 \right)^2 - \widetilde{x} \left( 10 \widetilde{x}^2 + 19 \widetilde{x} + 10 \right) \sigma^2 + \left( 8 \widetilde{x}^2 + 30 \widetilde{x} + 9 \right) \sigma ^4}{4 \sigma^4 \left( 4 \sigma^2 - 1 \right)^2 \left(\widetilde{x}^2 - \sigma^2\right) \left( \left( \widetilde{x} + 1 \right)^2 - \sigma^2\right)} \right] t^2 
   +\mathcal{O}\left(t^3\right)~,
\end{multline}
where we used for the sake of readability
\begin{equation}
\label{eq:ZDTypeIIInstantonReDefx}
    \widetilde{x} = \mathrm{i} x + \frac{1}{2}~.
\end{equation}
We checked \eqref{eq:ZDTypeIIInstanton} against \cite[eqs~A.3-A.13]{Sciarappa:2017hds} up to and including order $t^2$, and found perfect agreement\footnote{Looking at the product of the instanton parts of the Nekrasov \eqref{nek4d} and defect \eqref{def21} \eqref{eq:ZDTypeIIInstanton} partition functions, one finds equality with the ``chiral defect" \cite[eq~A.7]{Sciarappa:2017hds} if $\sigma_{\text{\cite{Sciarappa:2017hds}}} = - \epsilon \left(\widetilde{x} - 1 \right)$, and with the ``anti-chiral" defect \cite[eq~A.10]{Sciarappa:2017hds} if $\sigma_{\text{\cite{Sciarappa:2017hds}}} = \epsilon \left(\widetilde{x} - 1 \right)$. Note that since we are working in the self-dual phase of the $\Omega$-background, we have $\epsilon = \epsilon_1 = - \epsilon_2$.}.
On the other hand, one can see that the NS instanton part in \eqref{eq:ZInstantonF0} vanishes in the same four-dimensional limit.

In the 4d limit, the polynomial part \eqref{jp} reduces to
\begin{equation}
\label{eq:Zp4D}
    Z^\text{II,(p)}\left(x\right) = \lim_{g_s \to 0} Z^\text{(p)} \left( \exp\left( g_s x \right), g_s \right) = \exp\left( - \pi x \right) = \exp\left( - \pi \frac{y}{2 \epsilon} \right).
\end{equation}
Looking at the 1-loop part \eqref{eq:Z1LoopF0Defz} and \eqref{eq:Z1LoopF0} and the reparametrization \eqref{eq:ParametrizationDual4DLimit}, one can see that
\begin{equation}
    - \mathrm{e}^{2 \pi b^{-1} z_\pm} = \mathrm{e}^{- g_s \left(x \pm \ri \sigma\right)},
\end{equation}
and using the second integral representation of Faddeev's quantum dilogarithm \eqref{eq:FaddeevQuantumDilogarithmDoubleSineReps} gives
\begin{equation}
    \log \left( \Phi_b\left(z_\pm\right) \right) =
    - \mathrm{i} \frac{\mathrm{Li}_2 \left( \mathrm{e}^{- g_s \left(x \pm \ri \sigma\right)}\right)}{g_s}  - \mathrm{i} \int_0^{+\infty} \frac{\mathrm{d}t}{1 + \mathrm{e}^{2\pi t}} \log \left(\frac{1 - \mathrm{e}^{- g_s \left(x \pm \ri \sigma\right)} \mathrm{e}^{- g_s t}}{1 - \mathrm{e}^{- g_s \left(x \pm \ri \sigma\right)} \mathrm{e}^{g_s t}}\right).
\end{equation}
The expansion of for $g_s \to 0$ from above
\begin{equation}
\label{eq:4DlimitDiLog}
    - \mathrm{i} \frac{\mathrm{Li}_2 \left( \mathrm{e}^{- g_s \left(x \pm \ri \sigma\right)}\right)}{g_s} = - \mathrm{i} \frac{\pi^2}{6} \frac{1}{g_s} - \mathrm{i} \left( x \pm \ri \sigma \right) \log\left(g_s\right) - \mathrm{i} \left( x \pm \ri \sigma \right) \left[ \log \left( x \pm \ri \sigma \right) - 1 \right] + \mathcal{O}\left( g_s \right),     
\end{equation}
\begin{equation}
    - \mathrm{i} \int_0^{+\infty} \frac{\mathrm{d}t}{1 + \mathrm{e}^{2\pi t}} \log \left(\frac{1 - \mathrm{e}^{- g_s \left(x \pm \ri \sigma\right)} \mathrm{e}^{- g_s t}}{1 - \mathrm{e}^{- g_s \left(x \pm \ri \sigma\right)} \mathrm{e}^{g_s t}}\right) = - \mathrm{i} \int_0^{+\infty} \frac{\mathrm{d}t}{1 + \mathrm{e}^{2\pi t}} \log \left(\frac{x \pm \ri \sigma + t}{x \pm \ri \sigma - t}\right) + \mathcal{O}\left(g_s\right),
\end{equation}
The divergent terms in \eqref{eq:4DlimitDiLog} can be dealt with properly in the context of the open TS/ST correspondence \cite{matijn-alba}, but we will simply drop them.
By using the integral representation of the Gamma function, we get \cite[p.~8]{adamchik2003contributions}
\begin{multline}
\label{eq:ArcTanToLogGamma}
    - \mathrm{i} \int_0^{+\infty} \frac{\mathrm{d}t}{1 + \mathrm{e}^{2\pi t}} \log \left(\frac{x \pm \ri \sigma + t}{x \pm \ri \sigma - t}\right) = - 2 \int_0^{+\infty} \frac{\mathrm{d}t}{1 + \mathrm{e}^{2\pi t}} \arctan \left(\frac{t}{- \mathrm{i} \left(x \pm \ri \sigma\right)}\right)
    \\
    = \mathrm{i} \left(x \pm \ri \sigma\right)  \log \left( - \mathrm{i} \left(x \pm \ri \sigma\right) \right) - \mathrm{i} \left(x \pm \ri \sigma\right) - \frac{\log\left(2\pi\right)}{2} + \log\Gamma \left( - \mathrm{i} \left(x \pm \ri \sigma\right)  + \frac{1}{2} \right) \, ,
\end{multline}
where $\log \Gamma$ is the log Gamma function.
Hence, we get
\begin{equation}
    \lim_{g_s \to 0} \log \left(\Phi_b\left(z_\pm\right)\right) =  \log\Gamma \left( - \mathrm{i} \left(x \pm \ri \sigma\right)  + \frac{1}{2} \right) -  \frac{\log\left(2\pi\right)}{2} + \frac{\pi}{2}  \left( x \pm \ri \sigma \right) \, ,
\end{equation}
and combining with \eqref{eq:Zp4D} gives the perturbative part of the type II defect partition function
\begin{equation}
\label{eq:ZDTypeIIPertubative}
    \begin{split}
        Z^\text{II}_{\rm pert}\left( x, \sigma \right) &
        = \lim_{g_s \to 0} Z_{\mathbb{F}_0}^\text{(p)} \left( \exp\left( g_s x \right), g_s \right) Z_{\mathbb{F}_0}^\text{1-loop} \left( \exp\left( g_s x \right), \exp\left( 2 g_s \ri \sigma \right), g_s \right)
        \\
        & = \frac{\Gamma \left( - \mathrm{i} \left(x - \ri \sigma\right)  + \frac{1}{2} \right) \Gamma \left( - \mathrm{i} \left(x + \ri \sigma\right)  + \frac{1}{2} \right)}{2 \pi} \, .
    \end{split}
\end{equation}

The product of the perturbative \eqref{eq:ZDTypeIIPertubative} and instanton \eqref{eq:ZDTypeIIInstanton} \eqref{eq:ZDTypeIIInstantonReDefx} parts gives us then the complete partition function for the type II defect in 4d, $\mathcal{N} = 2$, $\mathrm{SU}\left(2\right)$ super Yang-Mills\footnote{We choose for the four-dimensional partition function a normalization that does not include the $1/2\pi$. In addition, the relation between eigenfunctions of the integral kernel \eqref{kernrho} and the defect partition function \eqref{eq:app:ZDTypeII} makes it convenient to include an extra factor $\exp \left( \mathrm{i} x \log \left( t \right) / 2 \right)$ in the latter.},
\begin{equation}
\label{eq:app:ZDTypeII}
    Z^\text{II}\left( x, t, \sigma \right)
    = \exp \left( \frac{\mathrm{i}}{2} x \log \left( t \right)\right)\Gamma \left( - \mathrm{i} x - \sigma + \frac{1}{2} \right) \Gamma \left( - \mathrm{i} x + \sigma + \frac{1}{2} \right) Z_{\rm inst}^\text{II}\left( x, t, \sigma \right)~.
\end{equation}
Note that we use a slightly different notation compared to the main text: what we call $x$ here is called $q$ elsewhere.

\section{From the matrix model identity to the eigenfunction identity}
\label{appendix:subsection:finaletofinale}

In this appendix, we argue for the equivalence between the identities \eqref{finalecompact} and \eqref{finale2}\footnote{ We will not be rigorous in switching the order of sums and integrals since all the quantities are conjecturally convergent.}. Our strategy is similar to the one used in the context of ABJM theory, see e.g.~\cite{Hatsuda:2012dt} and reference therein.

Let us start by writing \eqref{finalecompact} as
\begin{multline}
\label{tp1}
    \int_{ \IR+\ri \sigma_*}{\rd \sigma}  {\tan \left(2 \pi \sigma\right)\over\left(2 \cos (2 \pi  \sigma )\right)^N} \left(Z^{\rm II}_{\rm tot}\left( q, t, \sigma \right)+Z^{\rm II}_{\rm tot}\left( -q-{\ri \over 2}, t, \sigma+{1\over 2} \right)\right)
    \\
    =
    \mathrm{i} {2^{11/12}\sqrt{\pi}{ t^{3/16}  }\over \re^{3\zeta'(-1)}\re^{4 \sqrt{t}}(4\pi)^N}\int_{\mathbb{R}} {\rd x} \ \re^{-\ri {2} q x} \re^{- 4 t^{1/4} \cosh x + {x\over 2}}
\Psi_N(\re^{{x}}, t),
\end{multline}
where we have absorbed the $(-1)^N$ into a shift of $\sigma$, and $\sigma_* $ is a strictly positive number which guarantees that the integration contour on the left-hand side of \eqref{tp1} does not hit the poles of $Z^{\rm II}_{\rm tot}$. This is the case if $0 < \sigma_* < |{\rm Re}(q)|\neq 0.$ If ${\rm Re}(q)=0$, one can take $\sigma_*$ to be any strictly positive number as in \autoref{footnotess}.
For the sake of notation, let us define
\begin{align}
\label{g3}
    f(N) & = \ri {2^{11/12}\sqrt{\pi}{ t^{3/16}  }\over \re^{3\zeta'(-1)}\re^{4 \sqrt{t}}(4\pi)^N}\int_{\mathbb{R}} {\rd x} \
    \re^{-\ri {2} q x} \re^{- 4 t^{1/4} \cosh x+ {x\over 2}} \Psi_N(\re^{{x}}, t) \, ,
    \\
    g(\sigma) & =Z^{\rm II}_{\rm tot}\left(q, t, \sigma \right) + Z^{\rm II}_{\rm tot}\left( -q-{\ri \over 2}, t, \sigma + {1\over 2} \right) \, ,  
\end{align}
so that \eqref{tp1} becomes
\begin{multline}
\label{eq:FinaleExtraCompact}
    f\left(N\right) 
    = \int_\mathbb{R} \mathrm{d}\sigma \frac{\tan\left(2 \pi \left(\sigma + \mathrm{i} \sigma_* \right) \right)}{\left[ 2 \cos\left(2 \pi \left(\sigma + \mathrm{i} \sigma_* \right) \right)\right]^N} g\left(\sigma + \mathrm{i} \sigma_* \right)
    \\
    = \int_{-1/2}^{1/2} \mathrm{d} \sigma \frac{\tan\left(2 \pi \left(\sigma + \mathrm{i} \sigma_* \right) \right)}{\left[ 2 \cos\left(2 \pi \left(\sigma + \mathrm{i} \sigma_* \right) \right)\right]^N} \sum_{k \in \mathbb{Z}} g\left( \sigma + \mathrm{i} \sigma_* + k \right)~.
\end{multline}
Note that the second equality in \eqref{eq:FinaleExtraCompact} assumes some good analytic properties of $g$, for instance, $g$ is such that the sum over $n$ on the right-hand side of \eqref{eq:FinaleExtraCompact} is convergent. This is the case for \eqref{g3}. Furthermore, it is part of our conjecture that the function $\sum_{k \in \mathbb{Z}} g\left( \sigma + \mathrm{i} \sigma_* + k \right)$ is not only well defined but also an entire function of $\sigma$. Hence, we are free to deform the integration path in \eqref{eq:FinaleExtraCompact} to any path $\mathcal{C}_{ \left\{-1/2, 1/2\right\} }$, beginning at $\sigma = -1/2$ and ending at $\sigma = 1/2$, as long as we don't cross the poles coming from the tangent when $\sigma + \mathrm{i} \sigma_* \in \mathbb{Z} / 2 + 1 / 2$. Consider then the change of variables given by
\begin{equation}
\label{eq:mu(sigma)}
    \mu : \mathcal{C}_{ \left\{-\frac{1}{2}, \frac{1}{2} \right\} } \to \left] - \pi, \pi \right[: \sigma \mapsto \mu \left( \sigma \right) = - \mathrm{i} \log \left[ \frac{\cos \left( 2 \pi \left( \sigma + \mathrm{i} \sigma_* \right)\right)}{2 \pi}\right],
\end{equation}
which is well defined for a suitable choice of $\mathcal{C}_{ \left\{-1/2, 1/2\right\} }$. Then, we have for the integral that
\begin{equation}
\label{eq:FinaleExtraCompactRewritten}
    \frac{1}{2 \pi}\int_{- \pi}^{\pi} \mathrm{d} \mu \exp \left( - \mathrm{i} \mu N \right) \cdots
     = - \mathrm{i} \left( 2 \pi \right)^N \int_{-1/2}^{1/2} \mathrm{d} \sigma \frac{\tan\left(2 \pi \left(\sigma + \mathrm{i} \sigma_* \right) \right)}{\left[ \cos\left(2 \pi \left(\sigma + \mathrm{i} \sigma_* \right) \right)\right]^N} \cdots~,
\end{equation}
and hence, we can rewrite \eqref{eq:FinaleExtraCompact} as 
\begin{equation}
\label{eq:SomeIntermediateStep}
    - \mathrm{i} \left( 4 \pi \right)^N f \left( N \right) = \frac{1}{2 \pi} \int_{- \pi}^{\pi} \mathrm{d} \mu \exp \left( - \mathrm{i} \mu N \right) \sum_{k \in \mathbb{Z}}g\left( \sigma \left(\mu\right) + \mathrm{i} \sigma_* + k \right),
\end{equation}
where $\sigma \left( \mu \right)$ is the inverse of \eqref{eq:mu(sigma)}. 
Note that the integral on the right-hand side gives the $N$-th Fourier coefficient of the function $\sum_{k \in \mathbb{Z}} g\left( \sigma \left( \mu \right) + \mathrm{i} \sigma_* + k \right)$.
The full Fourier series leads then to
\begin{equation}
\label{eq:Finale2ExtraCompact}
    \begin{split}
        - \mathrm{i} \sum_{N \in \mathbb{N}} \left( 4 \pi \right)^N \exp\left(\mathrm{i} \nu N\right) f \left(N\right)
        & = \sum_{N \in \mathbb{Z}} \frac{\exp \left(\mathrm{i} \nu N \right)}{2 \pi} \int_{- \pi}^{\pi} \mathrm{d} \mu \exp \left( - \mathrm{i} \mu N \right)  \sum_{k \in \mathbb{Z}} g\left( \sigma \left(\mu\right) + \mathrm{i} \sigma_* + k \right)
        \\
        & =  \sum_{k \in \mathbb{Z}} g\left( \sigma \left(\nu \right) + \mathrm{i} \sigma_* + k \right).
    \end{split}
\end{equation}
where the last equality holds whenever the Fourier series on the previous line is convergent. We expect this to be true in our case even though we do not have a rigorous proof. 
We also used that $f(-N)=0$ for $N \in \IN \setminus \left\{ 0 \right\}$.

To go in the opposite direction from \eqref{eq:Finale2ExtraCompact} to \eqref{eq:FinaleExtraCompact}, one can apply Cauchy's residue theorem. Using the notation
\begin{equation}
    \kappa = \exp\left( \mathrm{i} \nu \right) = \frac{\cos\left( 2 \pi \left( \sigma + \mathrm{i} \sigma_* \right) \right)}{2 \pi}
\end{equation}
one can multiply both sides in \eqref{eq:Finale2ExtraCompact} by $1/ \kappa^{ N + 1}$ and integrate over $\kappa$ along an anticlockwise contour of radius 1 centered at the origin to arrive at
\begin{equation}
    \begin{split}
        \left( 2 \pi \right) \left( 4 \pi \right)^N f \left( N \right)
        & = \oint \frac{\mathrm{d} \kappa}{\kappa^{N+1}}  \sum_{k \in \mathbb{Z}} g\left( \sigma \left( - \mathrm{i} \log \left( \kappa \right) \right) + \mathrm{i} \sigma_* + k \right) 
        \\
        & = \mathrm{i} \int_{- \pi}^{\pi} \mathrm{d} \nu \exp \left( - \mathrm{i} \nu N \right) \sum_{k \in \mathbb{Z}} g\left( \sigma \left(\nu\right) + \mathrm{i} \sigma_* + k \right)
    \end{split}
\end{equation}
which is equivalent to \eqref{eq:SomeIntermediateStep} and hence by \eqref{eq:FinaleExtraCompactRewritten} to \eqref{eq:FinaleExtraCompact}. We can conclude that \eqref{eq:FinaleExtraCompact} and \eqref{eq:Finale2ExtraCompact} are indeed equivalent, provided $f$ and $g$ have good analytic properties which we assume to be the case.

\section{Elliptic integrals}
\label{app:EllipticIntegrals}

For the elliptic integrals, we follow the conventions of \textit{Wolfram Mathematica} \cite{reference.wolfram_2023_elliptick, reference.wolfram_2023_ellipticf, reference.wolfram_2023_elliptice, reference.wolfram_2023_ellipticpi}. The notation in \cite[pp.~8-10]{byrd1971handbook} is slightly different and we denote their elliptic integrals with a tilde. In particular, we have the normal or incomplete elliptic integral of the first kind for $k^2 \in \mathbb{R}$, $-\pi / 2 < \phi < \pi / 2$ and $k^2 \sin^2\left(\phi\right) < 1$, \cite{reference.wolfram_2023_ellipticf}, \cite[eq. 110.02]{byrd1971handbook},
\begin{equation}
\label{eq:EllipticF}
    \mathrm{F}\left( \phi  \middle| k^2 \right) = \int_0^\phi \frac{\mathrm{d} \theta}{\sqrt{1 - k^2 \sin^2\left(\theta\right)}} = \widetilde{F}\left( \phi, k \right),
\end{equation}
the normal or incomplete elliptic integral of the second kind for $k^2 < 1$ and $-\pi / 2 < \phi < \pi / 2$, \cite{reference.wolfram_2023_elliptice}, \cite[eq. 110.03]{byrd1971handbook},
\begin{equation}
\label{eq:EllipticE}
    \mathrm{E}\left(\phi  \middle| k^2 \right) = \int_0^\phi \mathrm{d} \theta \sqrt{1 - k^2 \sin^2\left(\theta\right)} = \widetilde{E}\left( \phi, k \right),
\end{equation}
and the normal or incomplete elliptic integral of the third kind for $k^2, \alpha^2 \in \mathbb{R}$, $-\pi / 2 < \phi < \pi / 2$ and $k^2 \sin^2\left(\phi\right) < 1$, \cite{reference.wolfram_2023_ellipticpi}, \cite[eq. 110.04]{byrd1971handbook},
\begin{equation}
\label{eq:EllipticPi}
    \Pi\left(\alpha^2; \phi \middle| k^2\right) = \int_0^\phi\frac{\mathrm{d} \theta}{\left(1-\alpha^2 \sin^2\left(\theta\right)\right)\sqrt{1 - k^2 \sin^2\left(\theta\right)}} =  \widetilde{\Pi}\left(\phi, \alpha^2, k\right).
\end{equation}
The complete elliptic integrals are obtained by taking $\phi = \pi / 2$,
\begin{equation}
\label{eq:CompleteEllipticIntegrals}
    \mathrm{K}\left(k^2\right) =\mathrm{F}\left( \frac{\pi}{2}  \middle| k^2 \right), \quad \qquad \mathrm{E}\left(k^2\right) =\mathrm{E}\left( \frac{\pi}{2}  \middle| k^2 \right), \quad \qquad \Pi\left(\alpha^2 \middle| k^2\right) =\Pi\left(\alpha^2; \frac{\pi}{2} \middle| k^2\right).
\end{equation}
The complete elliptic integrals of the first and second kind are analytic on $\mathbb{C}$ apart from a branch cut along the positive real line for $k^2 \geqslant 1$, and the complete elliptic integral of the third kind is analytic on $\mathbb{C}^2$ apart from similar branch cuts for $k^2, \alpha^2 \geqslant 1$ \cite{reference.wolfram_2023_elliptick, reference.wolfram_2023_elliptice, reference.wolfram_2023_ellipticpi}.

We also need a Jacobi elliptic function $\mathrm{sn}$ which is an inverse of the incomplete elliptic integral of the first kind \cite[p.~18]{byrd1971handbook},
\begin{equation}
\label{eq:JacobiSN}
    \mathrm{sn} \left( v \middle| k^2 \right) \, ,
    \qquad \qquad \qquad
    \mathrm{sn} \left( \mathrm{F}\left( \phi \middle| k^2 \right) \middle| k^2 \right) = \sin \left( \phi \right) \, ,
\end{equation}
which is sometimes called the sine amplitude.

\bibliographystyle{JHEP}

\bibliography{biblio}

\end{document}